# LOOKING FOR ENERGY IN DEMOGRAPHIC DATA — HOW TO DETECT SELF-ORGANIZATION IN HUMAN POPULATION DISTRIBUTION

Josep M Casas-Busquet   and   Agustí Poch-Parés

*Departament de Física. Universitat Politècnica de Catalunya. Diagonal 647. Barcelona. E08028. Catalunya. Spain.*

*Corresponding author. Agustí Poch-Parés*
*email: agusti.poch@upc.edu*

## Abstract

In the present work we study the relationship between population allocation and the combined effects of urban size and energy consumption, for two given areas and through a major part of the twentieth century. Along these lines a general application model is laid down which relates city-growth rates to initial inhabitants and to exosomatic energy increment, the deviations from it showing order in space and time—as shown in a series of maps which hint at unaccounted socioeconomic factors. The study of the maps by means of spectral analysis allows finding patterns which reinforce over time, in such a manner that spatial frequencies can be determined whose weight increases up so granting surface evolution estimation.

# Highlights

- The regressions can be considered as a way to quantifying the population's distribution
- Slopes correlate well with commercial primary energy increments to a 0.9 coefficient
- This result couples demography to energy and posits the emergence of self-order
- Our model's unexplained deviations show continuity in space
- The topography of the maps has been typified through the methodology exposed herein

# 1. Introduction

When studying natural systems the best approach to tackling the relationship between species is considering doing energy flow diagrams. This method plots trophic interactions between species and subsequent energy transfer between them, hence the study of patterns arising from this approach has set the rules of modern theoretical ecology. Energy flow diagrams can be seen through the lens of complex systems theory, especially when considering incidental appearance of self-organizing features—that is to say, some form of order arising out of the interactions. Generally speaking when a system is subjected to a flow of energy, self-organization arises to some degree. Prigogine proposed the so-called "dissipative structures" operating far from equilibrium in an environment with which they exchange energy and matter (Prigogine and Nicolis, 1971), the study of urban systems from this approach being a probed issue (Allen and Sanglier, 1978; Pulselli R.M. et al., 2005). Along the same vein Nicholas Georgescu-Roegen (1971) reintroduces the concept of exosomatic organ first presented in 1945 by Lotka, which is any tool necessary for the economic process. Georgescu underlines that exosomatic organs must be kept framed by a flow of matter and energy, which allows in due time for their unfolding and insertion in the species cultural evolution.

If we stick to energy, one of the main features of human species is its extensive use of auxiliary or exosomatic energy—in other words, energy not needed for its own nourishment. When we talk about energy used by man for travelling, heating, building, etc. we are referring to external, auxiliary or exosomatic energy. If the individual metabolism is considered to be about 150 watts, the amount of auxiliary energy used by humanity attains nowadays about seventeen times (IEA, 2014) the value necessary to feed the whole world. Obviously, the availability of auxiliary energy is not uniform around the planet, given that consumption in industrialized society reaches about one thousand times the amount needed for nourishing human population. Extending the early mentioned energy diagrams to account for the exosomatic point of view, Margalef (1997) suggests considering cities as species in a web whereby energy is transferred through. From this view alone one can imagine an energy distribution system featuring self-organizing processes so that human population may reorganize accordingly. In the present work we study the relationship between population redistribution and the combined effects of population density and energy consumption, regarding the latter as an incidental indicator of industrial fabric. It is worth mentioning that the relation we find implies only an energy-linked general trend on the zone, so that socio-economic issues apply when assessing single city's growth—as we shall see.

More precisely we have studied in the areas of Catalonia (32.000 km² in the NE of Spain) and Galicia (30.000 km² in the NW of Spain) the relationship between the municipalities' growth rates and their initial size, and how it connects to energy consumption. We also devote special attention to the case of the spatial regularities suggested by frequency-domain analysis of the maps which plot the deviations from the trend. We apply some concepts borrowed from open systems thermodynamics to the study of human population, which implemented in a fitted way reach quite complex results that suggest further research lines. Introducing energy in demographic studies is an innovative approach that despite having been nascently introduced in Catalan (Casas, 1989) as a partly published thesis, is original in English. Over and above initial proposal, the present study updates earlier work by extension of census, improves upon it by including validation through a second set of data, and broadens its scope by searching for structures through spectral analysis . All data processing from demographic census through energy correlations to maps and frequency-domain analysis, can be deemed a new technique whose singular results can be extended to a greater degree.

We use commercial primary energy values to quantifying exosomatic energy and a significant relationship is established between them and relative demographic growth, the latter expressed in terms of slopes of linear regressions. Along these lines a general application model is laid down which relates growth rates to initial population and to energy consumption, the deviations from it showing order in space and time—as shown in a series of maps which hint at unaccounted socioeconomic factors. The study of the maps allows finding regularities which reinforce over time, in such a manner that spatial frequencies can be determined whose weight increases up so granting surface evolution determination. The first idea leading to the unfolding of this work came from observing that roughly and sticking to 20th century, a town growth rate was a function of its initial population so that a correlation could be found between the latter two variables in municipalities of a given area. The second one was to suppose we were staring at the performance of transport network nodes given exosomatic energy pulses. The third idea was regarding the slope of the above correlation as a measure of the energy implied in the process. A fourth idea was to test the hypothesis that space was involved when analyzing deviations from the model thus providing a set of maps as further evidence. The fifth idea was to examine them via spectral analysis just to quantify incidental evolution.

## 2. Material and methods

The starting data set are the population censes of Catalonia and Galicia. Concerning Catalonia a database has been made from censes contained in

(Idescat, 2015) and (Iglèsies, 1961 and 1973). Regarding Galicia the original data source is (Ine, 2015).

The energy data sources are: (Ine, 2015), (M. Economía y Comercio, undated), (M. Instrucción Pública, 1917, 1918, 1922), (M. Trabajo y Estadística 1943, 1952), (Marín Quemada, 1978), (Ministerio de Industria y Energía, 1961, 1978), (PCM, 1934), and (PG, 1961).

The cartographic basis for plotting the maps has been produced via digitalization of originals supplied by Institut Cartogràfic de Catalunya and Xunta de Galicia. All data treatment was carried out with programs written in Python by the authors, and closing graphics and tables have been developed using Grapher and Surfer software packages.

As to the organization of the paper, discussion is made along with entering results hence the scheme "methods-results-discussion" iterates several times throughout.

## 3. Relating growth rates to initial population and exosomatic energy

We calculated the municipalities' growth rate through the model $P_t = P_o e^{kt}$, where $P_t$ is population at the moment t, $P_o$ is population at the moment 0, k is growth rate and t is time elapsed between 0 and t. This expression supposes the rate k to be independent from the initial population as shown in the differential equation $dP/dt = kP$, whose solution it is. This model's use justifies by its simplicity as long as the time periods considered are not too extended, because in the long run culture-related positive deviations can appear which are the type of phenomena we are about to quantify. Once the values of k for a certain area and time are given, we study whether they get influenced by the initial population— this we do by means of regression of k on ln(P) where ln is the Napierian logarithm.

We calculated the value of k for all the area of Catalonia ( 931 municipalities) through the years 1717 to 2013 , and as a validation data set the same values were calculated for Galicia (321 municipalities)  over 1900 to 2014. For the sake of precision, our model turns

$$P(m,d+1) = P(m,d)e^{k(m,d)t} \qquad (1)$$

to calculate k(m,d) for each municipality (m) and date (d) and so the above introduced k will be referenced thereafter.  As stated, we probe the influence of the initial population on the growth rate values by way of regressing  k(m,d) upon ln(P(m,d)).  We have thus obtained regression lines for all considered time spans so that each follows an equation of  type:

$$k_M(m,d) = A(d) + B(d)\ln(P(m,d)) \qquad (2)$$

where $k_M(m,d)$  is growth rate value matching the regression, B(d) stands for slope of the line, ln(P(m,d) is population's Napierian logarithm regarding municipality m and date d, and A(d) is independent term.

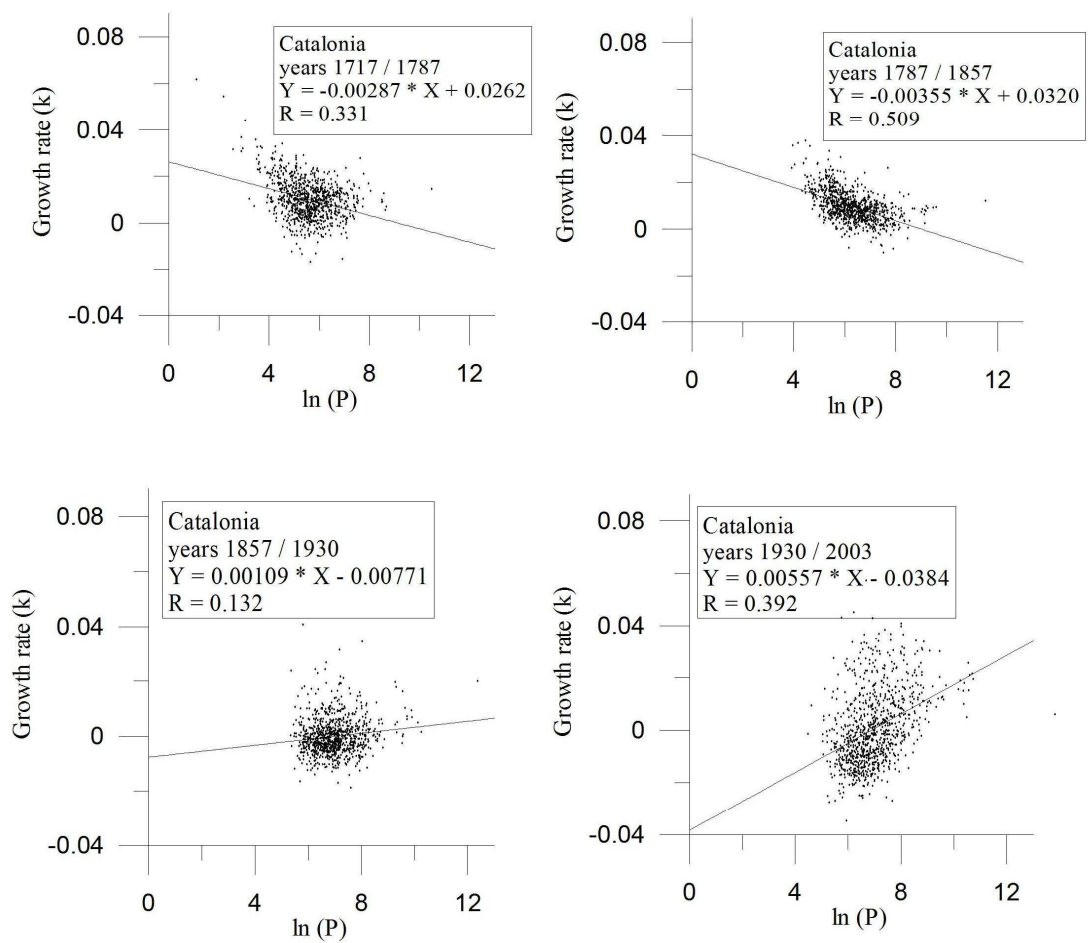

Figure 1. *Influence along time of initial population over growth rate, in Catalonia. Given the growth rates —or values of k— for a certain period, we study whether they get influenced by the initial population through regression of k on ln(P), which is the Napierian logarithm of mentioned population. Each diagram plots the behaviour of the whole set of municipalities through the referred interval. Note the switch of the slope across time.*

We made a first search on Catalonia considering (approximate) seventy-year periods beginning on 1717; the approximation lies in the fact of relying upon the data in the censal survey, which collected in an uneven pattern produces somewhat irregular periods—hence we get Figure 1. Case (1717/1787) shows a trend towards decreasing growth as population increases, this being more acute regarding period (1787/1857). Still, correlation is bigger in the latter case thus pointing to a more accentuated behaviour. The trend reverses in period (1857/1930) starting to display a bigger growth of major cities, this becoming more clear during case (1930/2003) where big populations appear plainly favored. The time series of the regression slopes can be seen in Figure 2; remarkably, we can see the three last points nearly aligning each other thus displaying a trend which once considered suggests a further investigation into the data concerning the 20th century as a clear exponent of the former drift. We state that from 1857, big cities grow faster—so we likewise propose ourselves analyzing the correlation between this tendency and the consumption of exosomatic energy, inspired by the likelihood that the latter outlines the settlement of industrial frame which would, at his time create the conditions for a demographic change to occur. Additionally, we also propose interpreting the alleged tendency through the prism of self-organizing phenomena.

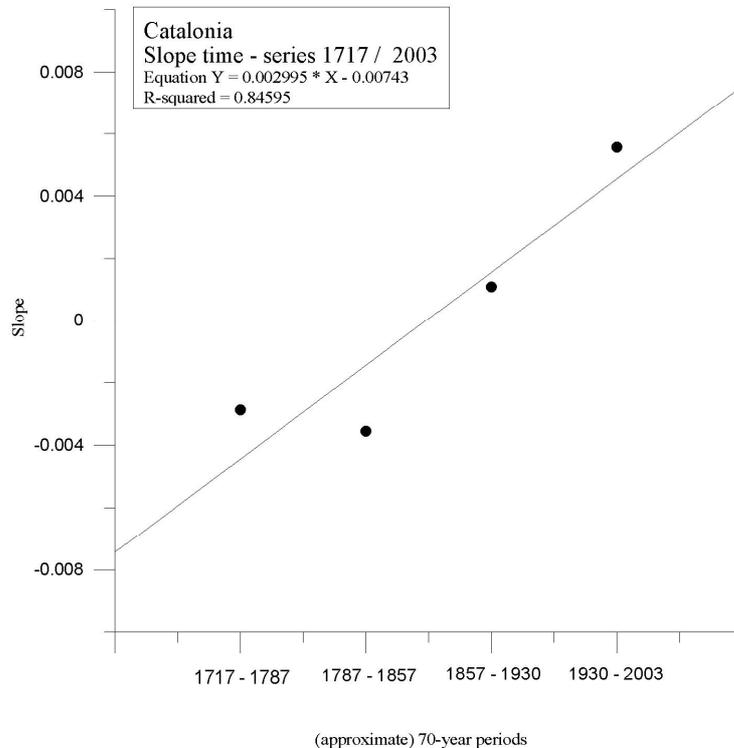

Figure 2. *Trend reversal. From 1857 onwards, big cities grow faster.*

*That's why the slopes turn positive disclosing that in this extent of human development, patterns arise which entail heightened returns, chiefly favoring big hubs. The values on the Y-axis are the slopes obtained in Fig. 1, and X- axis represents the (approximate) 70-year periods considered.*

The adduced research firstly yields Table 1, showing the decadal evolution of slopes belonging to Catalonia and Galicia for the best suited span over the 20th century, calculated in the same way as aforesaid and whose inclination values are comparable to those from above-cited periods 1857/1930 and 1930/2003. We will shortly be discussing the graphical plotting of these values which as a whole, give a sample of the statement that—granted the positive slopes— the value of k is not independent from initial population but just a function of it, getting bigger along with urban size. Thus, in this stretch of cultural evolution structures are created which imply amplified outcomes, i.e. self-organizing features, especially showing in big cities. If we were to include those in our differential equation we should write: dP/dt = k(P)P indicating the dependence of k on initial population.

| Table 1 Coefficients of regression lines [a] | | | |
|---|---|---|---|
| | A(d) | B(d) | Correlation (R) |
| **Catalonia** | | | |
| 1900-1910 | -5,341E-03 | 1,196E-03 | 9,469E-02 |
| 1910-1920 | -5,006E-03 | 1,340E-03 | 9,060E-02 |
| 1920-1930 | -3,845E-02 | 5,594E-03 | 2,667E-01 |
| 1930-1940 | -2,718E-02 | 3,053E-03 | 2,288E-01 |
| 1940-1950 | -2,326E-02 | 3,458E-03 | 2,584E-01 |
| 1950-1960 | -6,926E-02 | 9,941E-03 | 3,748E-01 |
| 1960-1970 | -3,582E-01 | 4,619E-02 | 3,583E-01 |
| **Galicia** | | | |
| 1900-1910 | -9,553E-03 | 1,441E-03 | 1,358E-01 |
| 1910-1920 | -1,077E-02 | 1,423E-03 | 1,278E-01 |
| 1920-1930 | -2,298E-02 | 2,959E-03 | 2,382E-01 |
| 1930-1940 | -2,443E-02 | 3,690E-03 | 2,586E-01 |
| 1940-1950 | -3,116E-02 | 3,600E-03 | 2,362E-01 |
| 1950-1960 | -3,518E-02 | 3,522E-03 | 2,368E-01 |
| 1960-1970 | -7,166E-02 | 7,179E-03 | 3,285E-01 |
| 1970-1981 | -7,438E-02 | 8,128E-03 | 4,037E-01 |

(a) A(d) is the independent term of regression line, and B(d) the slope as detailed in formula (2) above

Similar approaches consist in the equation  $dP/dt = kP^w$, where w stands for self-organization, though sometimes called autocatalysis (Peschel and Mende, 1986).  Hyperbolic world population models follow this equation like the one by Kapitsa (1992, 2014) which includes stabilization parameters leading to population saturation.  In fact our starting model is a particular instance of this equation namely when w=1, so reaching exponential growth.  The dependence of rate k on population can be compared with the so-called Gibrat's law (1931). Gibrat was a French engineer who lived between 1904 and 1980 and developed a statistical law whereby there is no relation between a city's population and its growth. There's a lot of controversy on the subject and authors can be found providing evidence for and against. Our results are in opposition to Gibrat's law since they generally show growth being a function of population.

Our regressions can be considered as a way to quantifying the population's horizontal distribution, that is to say the way in which it clusters in bigger or smaller centres. Correlation coefficient mainly increases over the 20th century (cf. table 1), and this can be interpreted as an unification in the behavior of the area as a reply to continuity and gain in energy use. Some hydrodynamical and applied physics models connect  regression slopes to implied energy, perhaps inspired by the work of Kolmogoroff  (1941) which posits a formula for the distribution of energy among turbulence hence delivering a fixed slope.  Given that our regressions quantify population clustering in knots of a transport network, we studied the relationship between their slopes and exosomatic energy supply, on trust that it would reflect some dependence of the self-organising level in the area. We recall here the introduced concept of dissipative structures whereby exosomatic energy ingress prompts appearance of organization, this  outlined  in our case by the slopes which quantify population spread—even if the sort of organization we are remarking herein belongs to a statistics-based method, thus abstracting away from the usual physical systems and settling the former at the verge of the theory.

We therefore performed a second regression (cf. Figure 3 for Catalonia and Figure 4 for Galicia), and found that on a logarithmic scale,  slopes correlate linearly well with commercial primary energy increments to a 0.9 coefficient in both areas. Put another way, and for the case of Catalonia, the slope nearly scales with energy as [ΔE(d)]$^{1.0485}$/antilogarithm 9.3316 with ΔE(d) standing for increase in energy consumption in date (d), and bearing in mind that 1.0485 is the line's slope in Figure 3 (actually the incline of the new correlation), and 9.3316 its independent term.

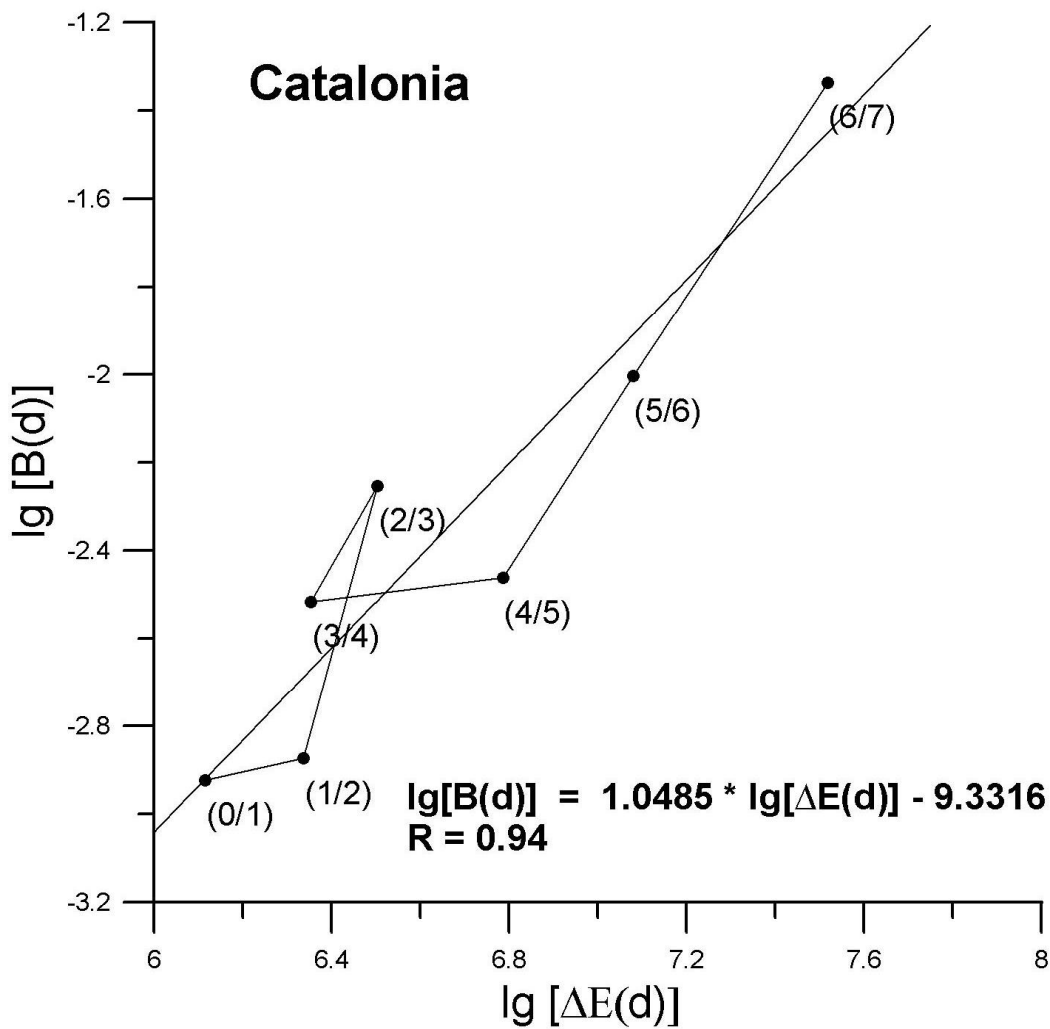

Figure 3. *Linear correlation of slopes with commercial primary energy increments for Catalonia, on a logarithmic scale. During a long stretch of 20th century, the measured relationship between the municipalities' growth rates and their initial size correlates highly with commercial primary energy to a 0.9 coefficient. The values on the Y-axis are the*

*logarithms of the slopes of the lines in Table 1, while X-axis features logarithms of commercial primary energy increments expressed in tce.*

It is as if for every energy increase there was a given population spread, or as if the former defined a migratory flow pattern. This being mindful that slopes just define trends since correlations between k and ln(P) are low though improving through time, and that regressions do not take space in account which shows being an important element as we will shortly see. It must also be noted that these slopes reflect the drive of roughly 1000 regression points thus adding to the reliance of the approach. In view of results and back to equation (2) we can nearly interchange slope B(d) by energy and hypothesize that it is it together with logarithm of population that accounts for a good share of actual demographic distribution though socioeconomic factors will also be shown to influence. This is a major result as it closely couples demography to energy and posits the emergence of self-order as a consequence of exosomatic energy consumption.

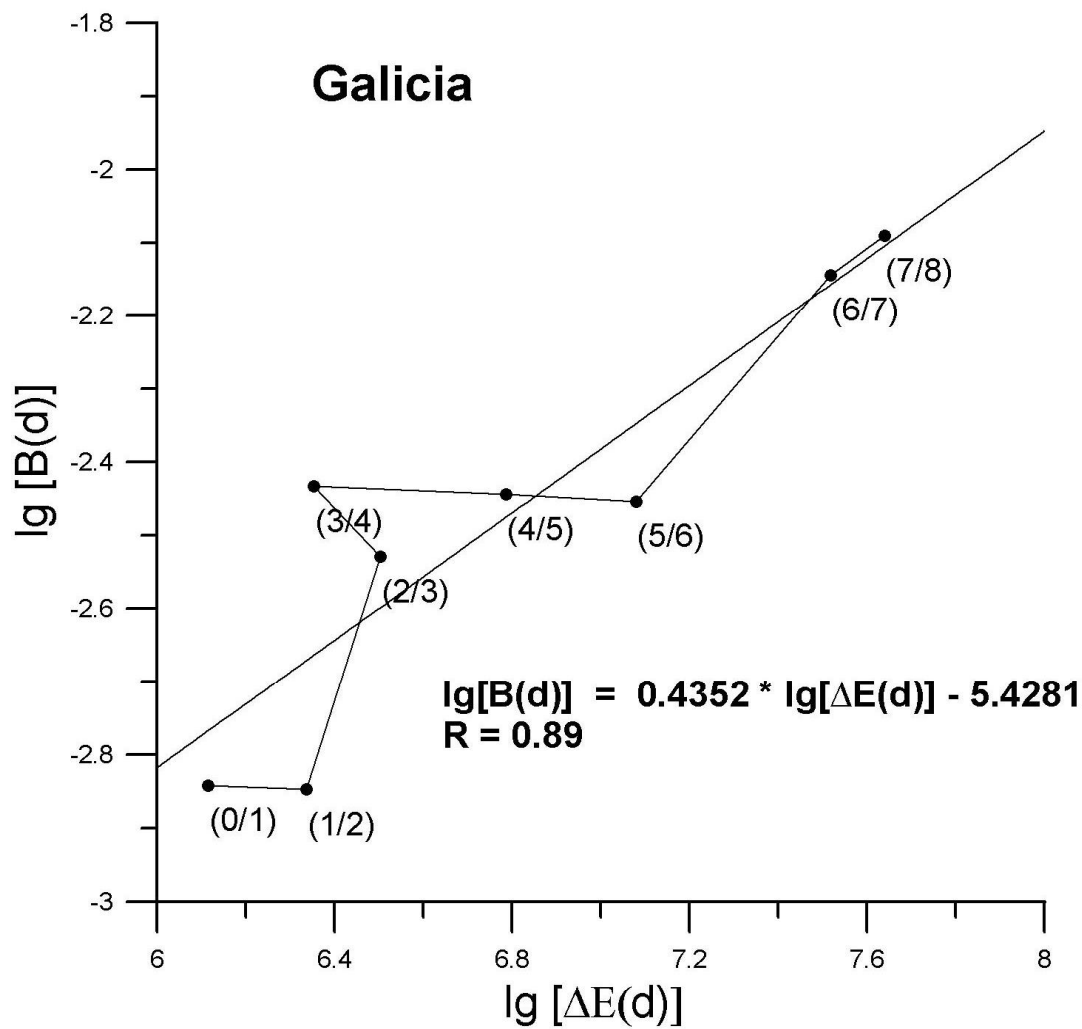

Figure 4. T*he case of Galicia as regards the correlation of slopes with energy increments, on a logarithmic scale, until year 1981. The values on both axis emulate those in Figure 3.*

Some important questions arise here, like whether has the issue been addressed previously, or whether does the correlation between energy and slopes induce causality, or how long in time this phenomenon is going to last. With regards to the first one we know of no similar work so far, though if considering population distribution to be an aspect of economic activity some authors tend to relate the latter to energy consumption (Hall *et al.*, 1986; Hall and King, 2011) even though if there is a long shot that our approach be included. Regarding second question, one has to agree that empirical observed correlation is a necessary but not sufficient condition for causality, which can only be defended when variables relate by means of diagrams or causal analysis (Pearl 2009). In our case we should be able to shape an energy circuit in style of H.T. Odum—see (Odum and Peterson, 1996)—and experiment all the way through. In present work we only introduce and discuss the correlations obtained, expecting that performing these calculations in other areas of the world will allow validation of results and improvement of the method.

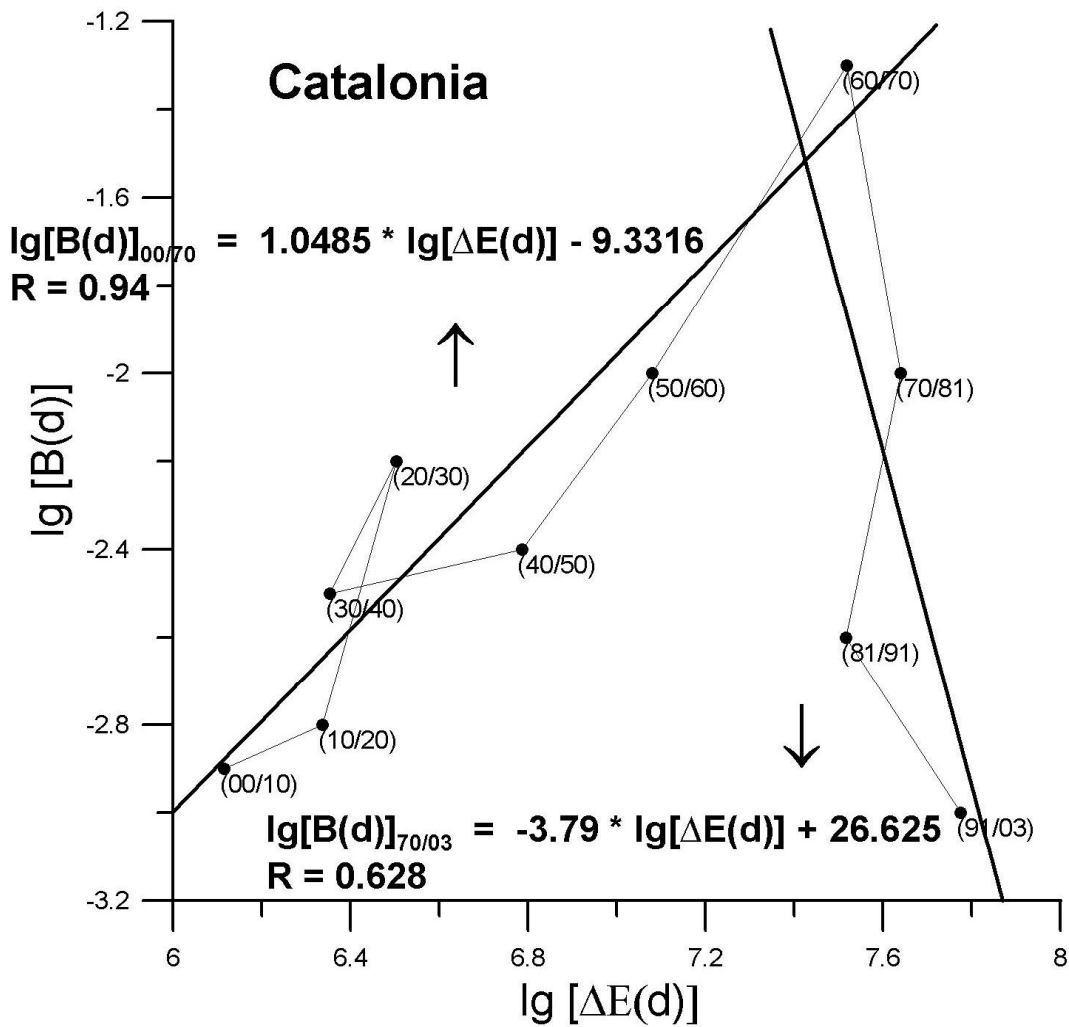

Figure 5. *Updated correlation of slopes with energy increments in the case of Catalonia, on a logarithmic scale, stretching to year 2003. There is a clear downturn so that the last counts of the epoch seem to follow another dynamics as shown by a second adjusted line . The values on both axis emulate those in Figure 3.*

As for the third question we further extended the analysis to recent years and reached extinction for Catalonia and maintenance for Galicia as follows. Stretching the range to year 2003, in the case of Catalonia there is a marked decline so that the last points of the century seem to follow yet another trend (see Fig. 5). The behaviour of the area can be interpreted as reaching a saturation point, as commonly seen in the biggest cities when considering their growth as a function of initial population. Here the whole zone seems to access into a further step wherein small populations efficiently compete against larger opponents in virtue of a previously generated pattern. As regards Galicia, notwithstanding the possibility of a new trend defined by the last decade, the system seems to keep within limits (Fig. 6) pointing towards a range of settled futures whereby this less developed rural area eventually reaches comparable transition.

However, this model's validity is limited to the extent required for the correlation to be applicable, i.e. in case of a sharp energy shortage like the one expectable during a lasting crisis, the low value of lg[ΔE(d)] — logarithm of commercial primary energy increment—would place its associated point well apart from the regression line thus opening yet another sequence. In fact, that is what we observed when data from decade 2003/2013 were considered (see Figures 7 and 8). In both gauged areas the corresponding point is scattered away from the main cloud of dots, even if the slope values still keep their marks over such shortfalls—hence allowing for future attuned developments. It's worth noting that in order to plot the negative energy value, logarithmic transformation of x-axis has been avoided; therefore, when comparing with former figures, bigger values should be imagined to be closer to the cluster as a corrective measure. Thus, back to the discussion, present-day data produce a standing-by position which is original enough, in that it exhibits the consequences of the first reversal in energy increment since the early 20th century. Concerning the values used to appraise energy (cf. Section 2), they belong to commercial primary energy in Spain, assuming that the consumption in the studied areas has always been a steady fraction of the total. The good correlation obtained in most of the cases emphasizes the trend which is embedded into those sets of energy data.

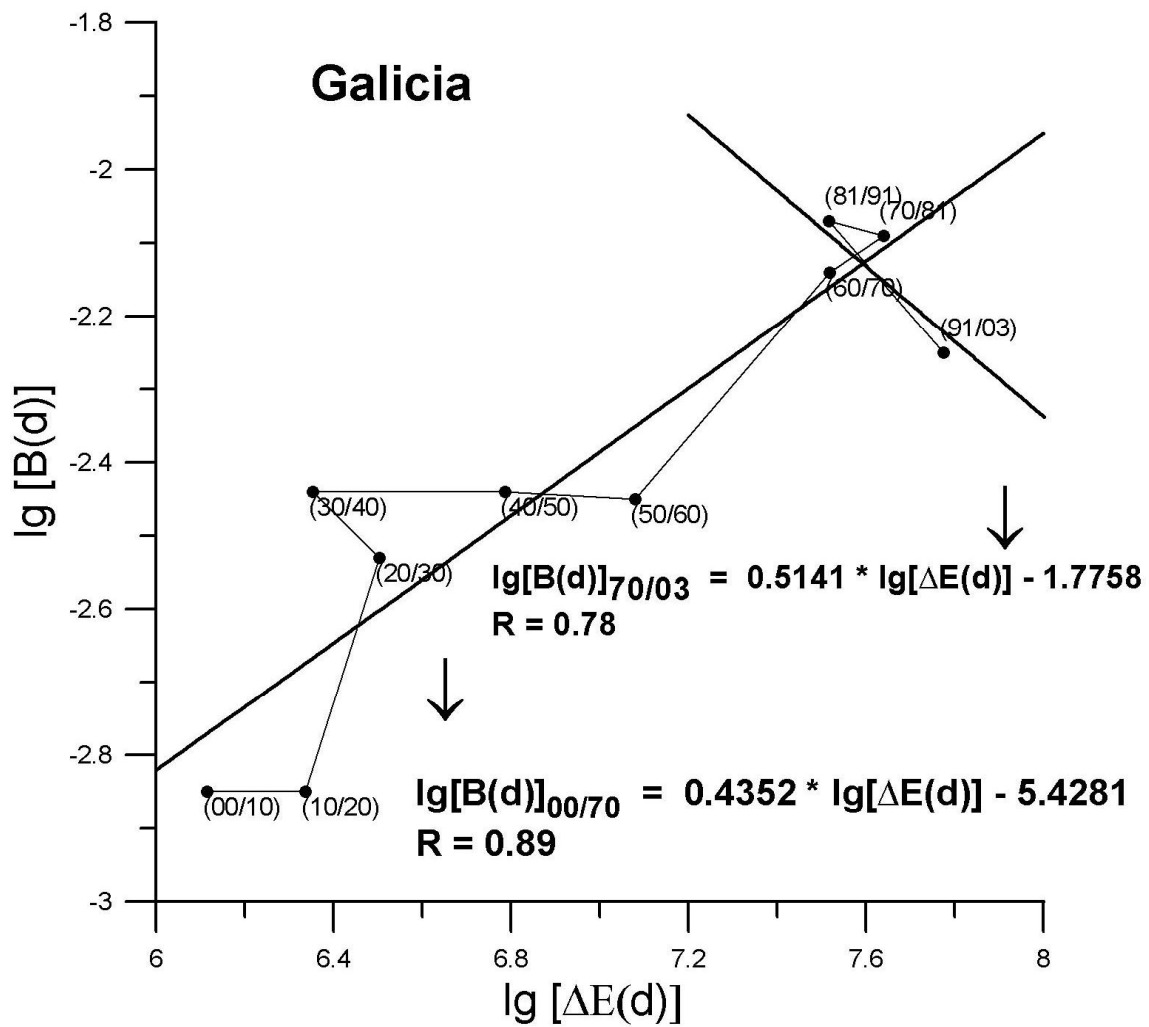

Figure 6. *Revised correlation of slopes with energy increments as regards Galicia, on a logarithmic scale, stretching to year 2003. In the main, slopes keep the trend in this less grown zone, even if the last decade could define a new dynamics. The values on both axis emulate those in Figure 3.*

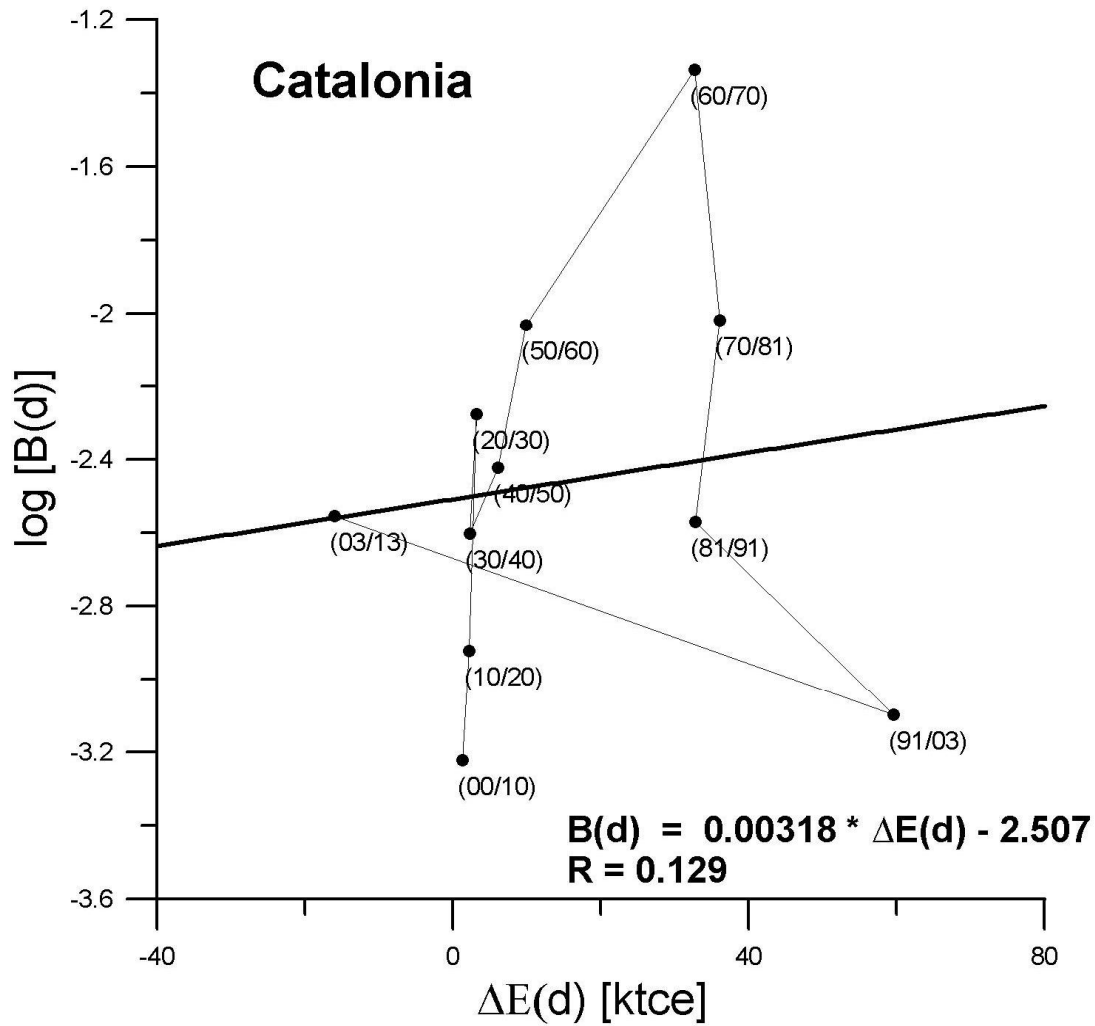

Figure 7. a) *First reversal in energy increment since the early 20th century. Correlation of slopes with energy as regards Catalonia—without logarithmic transformation of X-axis in order to plot the negative energy value respecting decade 2003/2013; note accordingly that whenever comparing with former figures, bigger values should be imagined to be more attached to the group. The values on the Y-axis are logarithms of the (updated) slopes conforming to lines in Table 1, while X-axis features commercial primary energy increments expressed in ktce;*

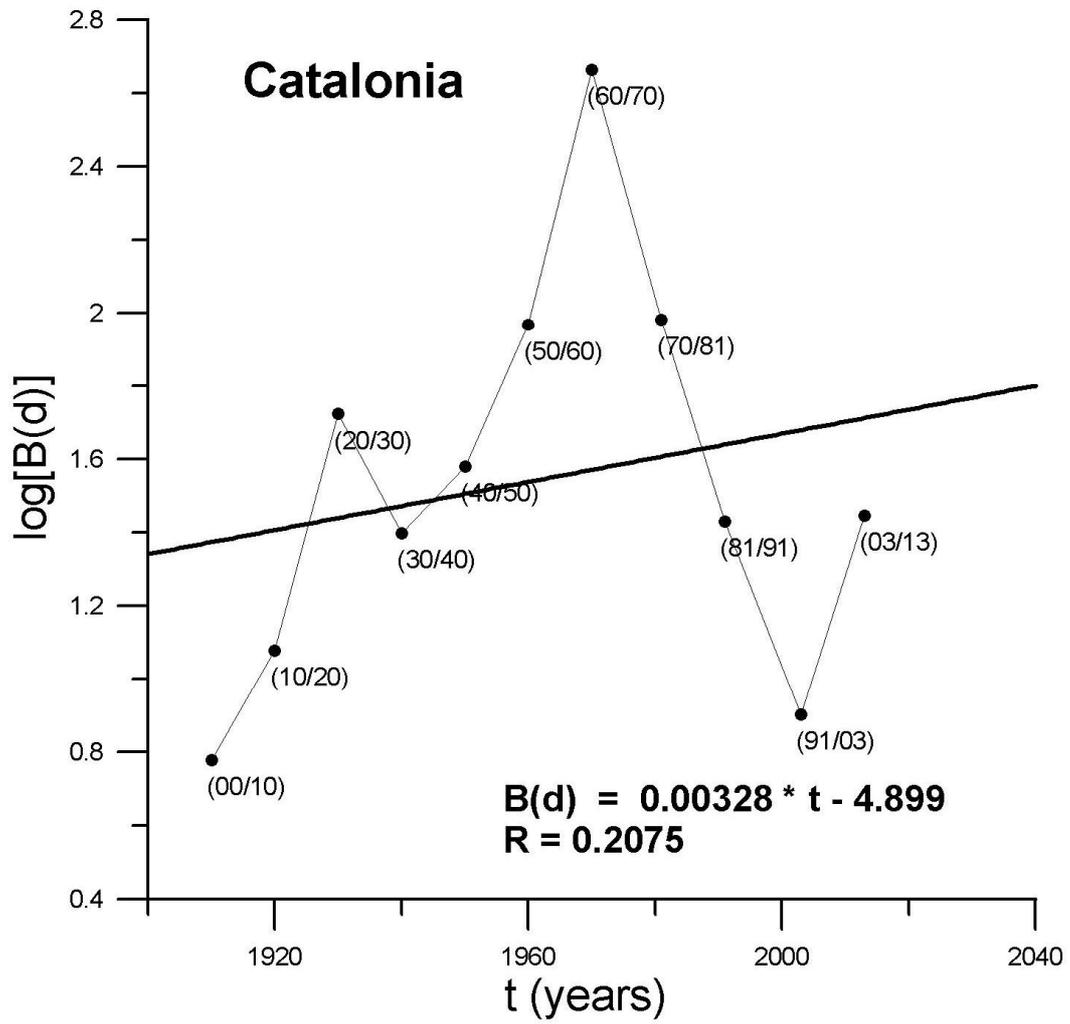

Figure 7. b) *Time-series of logarithms of slopes through the 20th century, in the case of Catalonia*

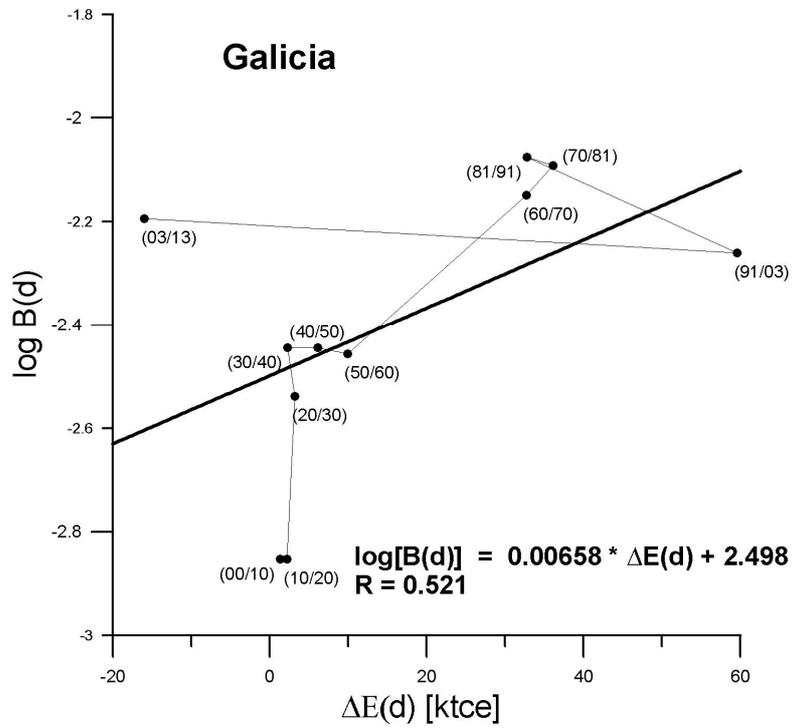

Figure 8. a) *Result of first reversal in energy increment as for Galicia. Data without logarithmic transformation regarding X-axis, so annotations in Figure 7 apply*;

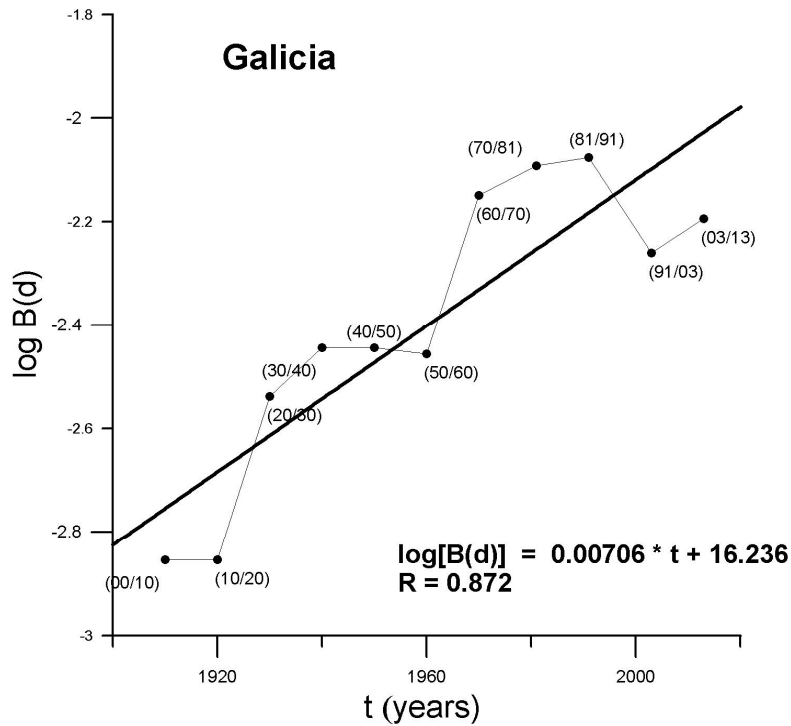

Figure 8. b) *Time-series of logarithms of slopes through the 20th century, in the case of Galicia.*

## 4. Spatial meaning of correlation deviations

Regressions of k on ln(N0) do not show such a good correlation. The problem often lies in some sort of structure which remains hidden until space is kept in view. To apply this approach to our data we consider two values of growth rate for each municipality, a real one calculated through model (1) which we name k(m,d) and a theoretical one obtained after the model (2) and termed $k_M$ (see Fig.9). Residual δk(m,d) will be the difference:

$$\delta k(m,d) = k(m,d) - k_M(m,d) \qquad (3)$$

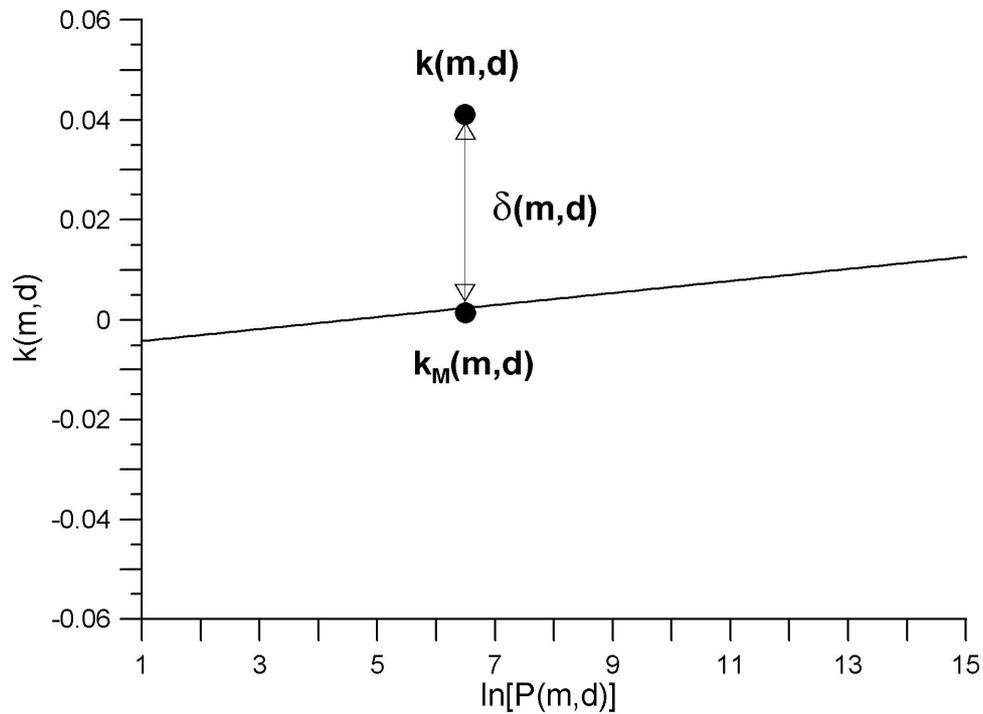

Figure 9. *A geometrical view of residual as defined in equation (3). A positive value of δk(m,d) implies that a given municipality scores above the trend line, according to its inhabitants. The values on both axis emulate those in Figure 1.*

For a given decade we plot values δk(m,d) on a map, thus getting a spatial design of our model's unexplained deviations. These show continuity in space, easy to sketch by contour lines, deploying a structure not detected by means of regression alone. Those forms can be interpreted as the unfolding in space of socio-economic factors not explained by the model and may be subject to further review. When considering positive outcomes, special population-clustering zones are detected, therefore emphasizing outstanding transport domains. We can observe in Figures 10 and 11 several sample maps for Catalonia and Galicia respectively.

Figure 10. a)

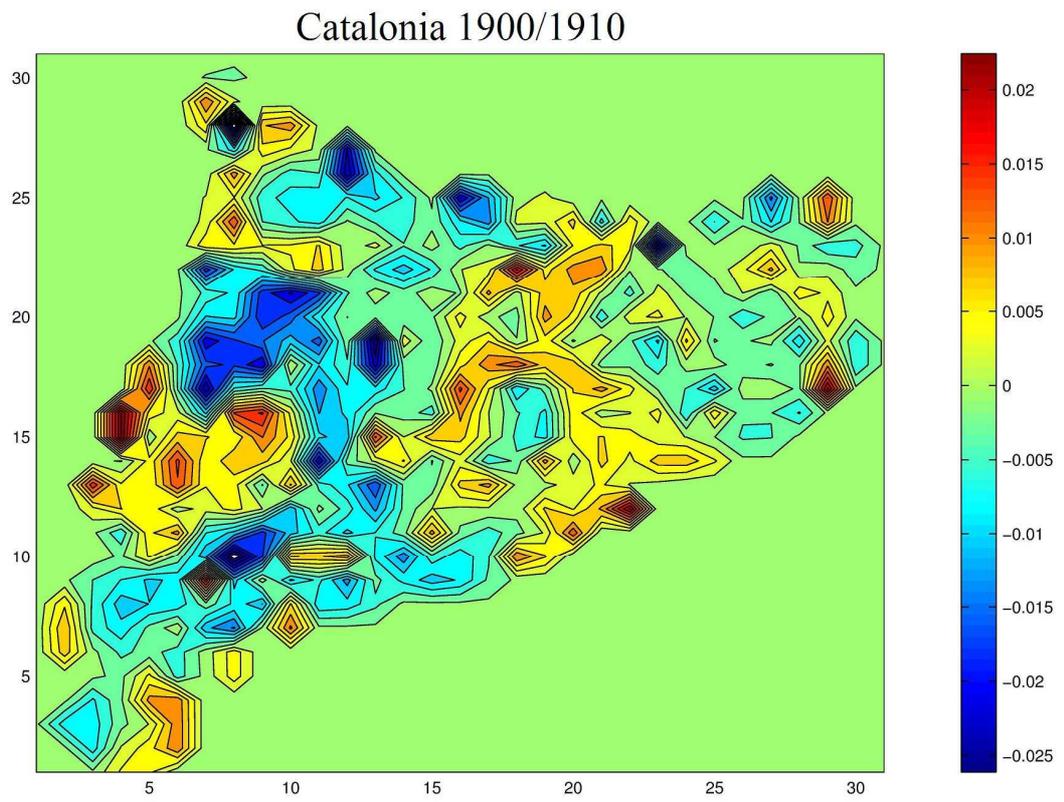

Figure 10. b)

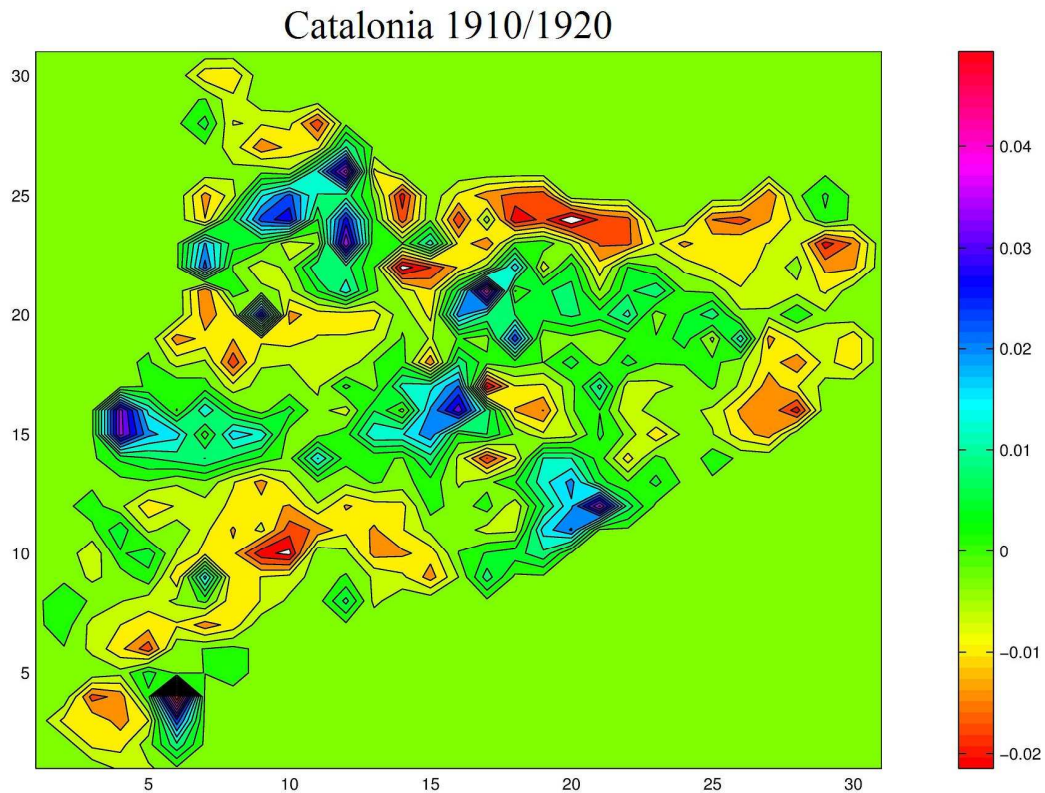

Figure 10. *Deploying of a structure not detected by means of regression alone. For a given decade, values δk(m,d) are plotted on a map, thus getting a spatial design of the model's unexplained deviations. These show continuity in space, easy to sketch by contour lines. The cases of Catalonia during decades (1900/1910) —Figure a)— and (1910/1920) —Figure b)— are depicted. The shading of maps is unrelated—each map displays specific shading according to its attached gradation scale.*

When considered, studying the maps over time comes up with better ordered reliefs, that is, profiles where different types of municipality appear less intermixed. From our approach, and causal analysis lacking, we can discuss this result in saying that the period covered in present work comprises the transition of a territory from a situation with more diversity and less connectivity, to a stage in which the intensification of energy consumption has induced less diversity and more interdependence. Catalonia—and Galicia to a lesser degree—would have passed from a situation in which remote municipalities were less dependent on each other, to an opposite one in which subordinate districts are the basis for larger and steady areas owing to an ever-increasing exosomatic metabolism.

Figure 11. a)

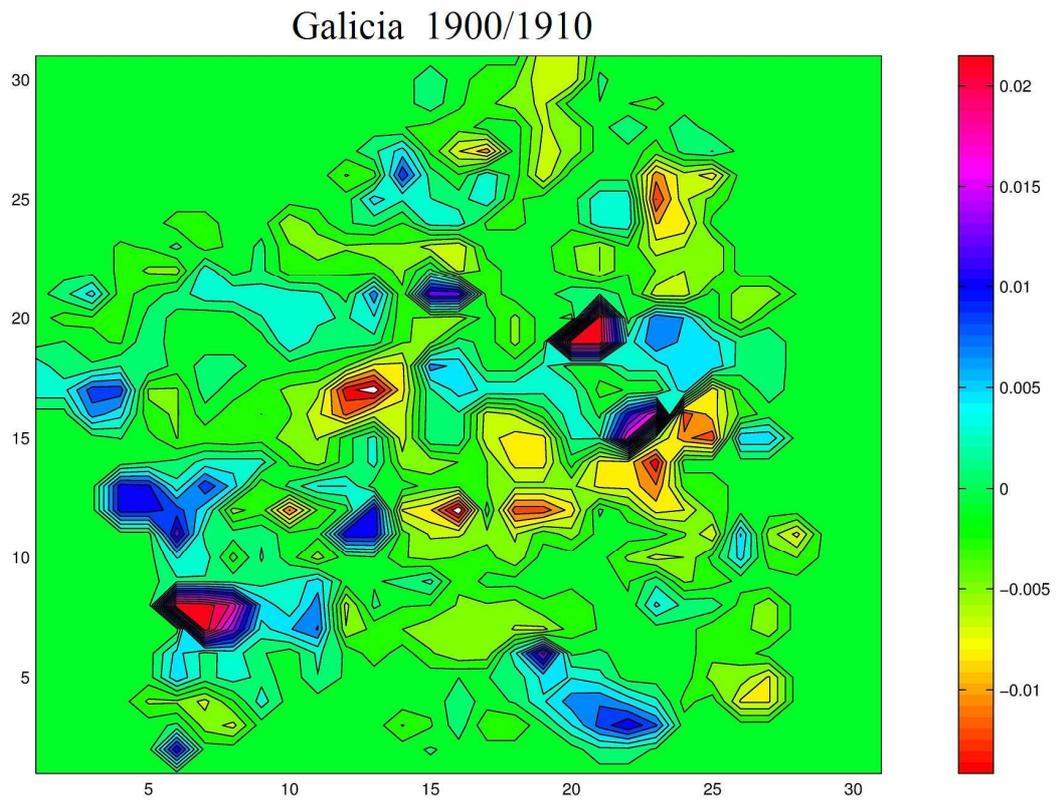

Figure 11. b)

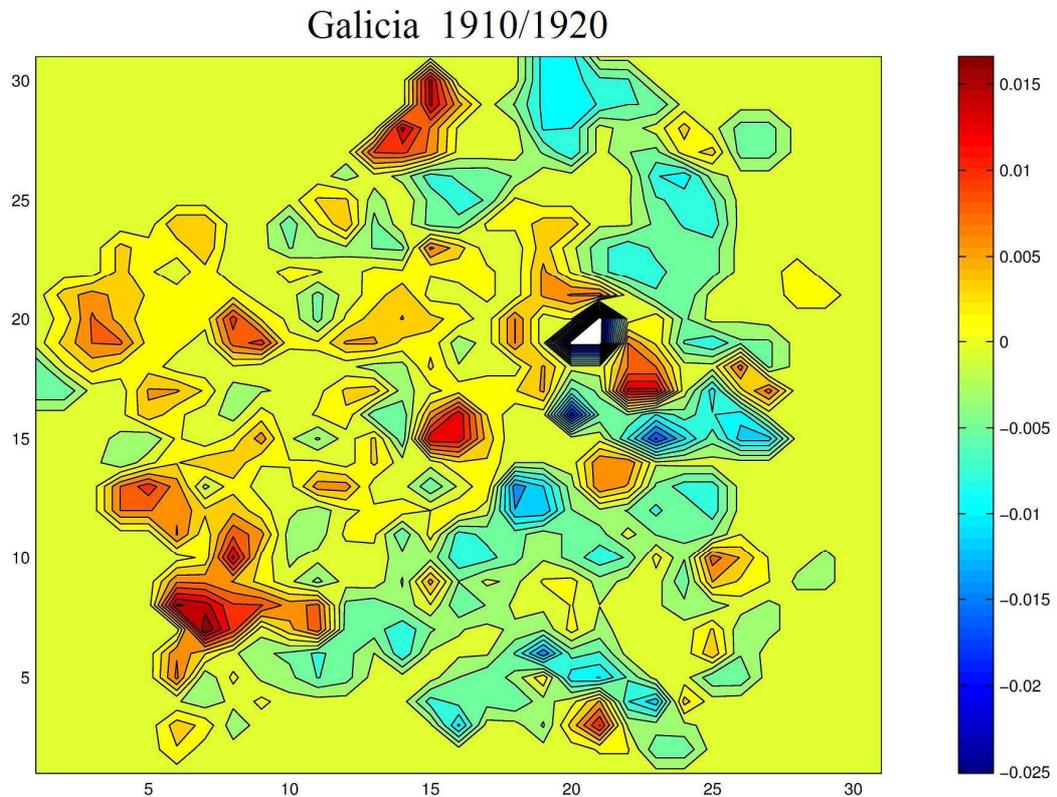

Figure 11. *Unfolding of socio-economic factors not explained by the model regarding Galicia during decades (1900/1910) —Figure a)— and (1910/1920) —Figure b)—. The values in all axis emulate those in Figure 10. Shading of maps unrelated.*

Further information from the maps can be extracted by means of merely adding them together on a point by point basis. By this doing we highlight outstanding centers, namely the ones spanning over the mean most of the time. These organize in space still, as shown in Figure 12 for decadal maps between 1900 and 1970 concerning Catalonia. Yet, given a series of maps, one can wonder whether is it possible to gauge evolution—that is, detecting some sort of phenomenon unfolding through them besides simply pooling the points. This we will meet in next section.

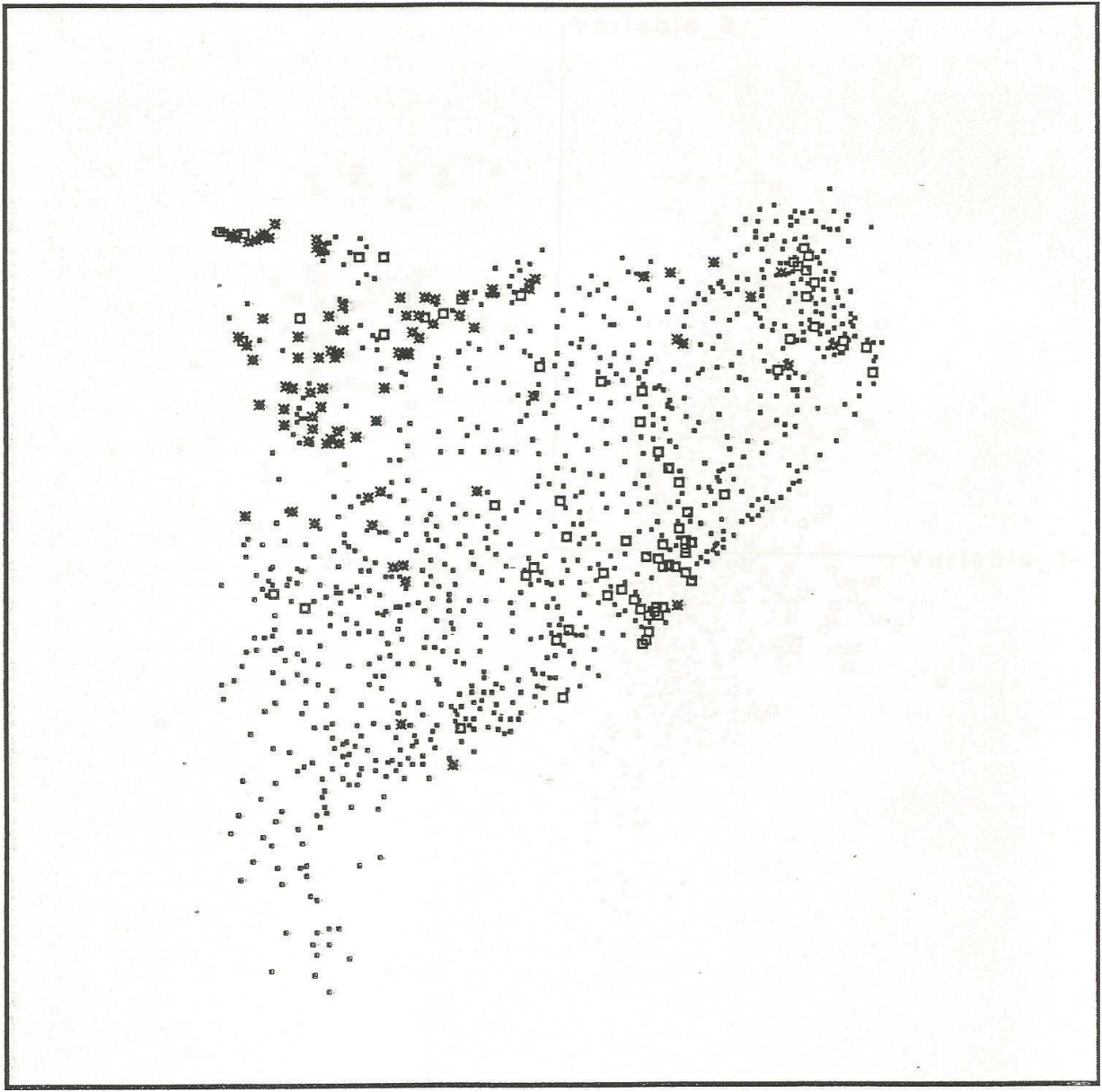

Figure 12. *A display of the addition of decadal maps between 1900 and 1970 involving Catalonia. Adding the maps together on a point by point basis can extract further information—thus highlighting areas spanning over the mean most of the time, composed here by squares featuring municipalities which reach maximum values. Centers with minimal values show asterisks.*

## 5. Spectral analysis of the maps

## 5.1 The Fourier transform

Starting from a curve like one plotting temperature against time, the Fourier Transform allows producing a series of regular curves that add up to the original one. In this way we uncover the subjacent regularities of the curve which we term "frequencies", each with an associated weight i.e. relative importance. This information we call "spectrum" which is usually expressed as a plotting of weight against frequency. Essentially, performing Fourier Transform is filtering every point in a set in search for the amount it contributes to the weight of a given frequency, and this iterated through all the spectrum of frequencies.

The Fourier Transform and its counterpart the Discrete Fourier Transform (DFT), are widely used today in such different fields as supercomputing to quantum mechanics. DFT applies when we deal with discretely sampled signals, like the scattered points we have on a map, when just a finite number of frequencies is computed. The Fourier Transform can be executed over signals propagating through different dimensional spaces; that is, we can analyze from one-dimensional curves like time series to n-dimensional volumes in Neural Networks. In our case we will study surfaces hence we will work with two dimensions. Following Fourier, we know that every map can be splitted —see Fig. 13— into a collection —which we call spectrum— of regular surfaces —or frequencies—.

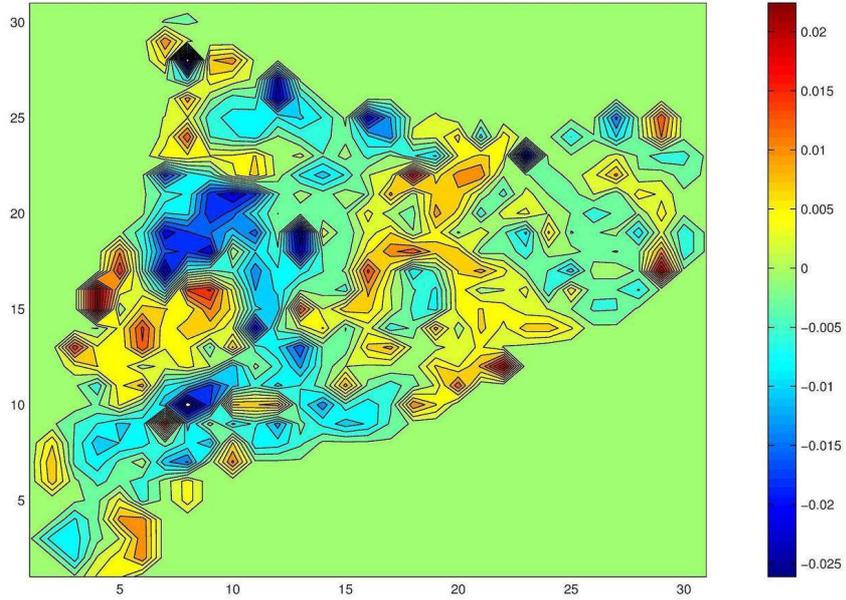

Fourier Transform

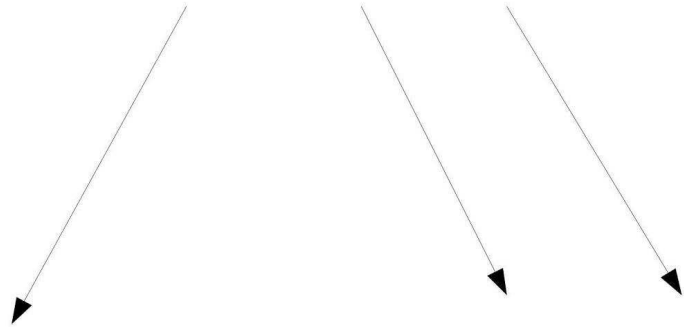 +  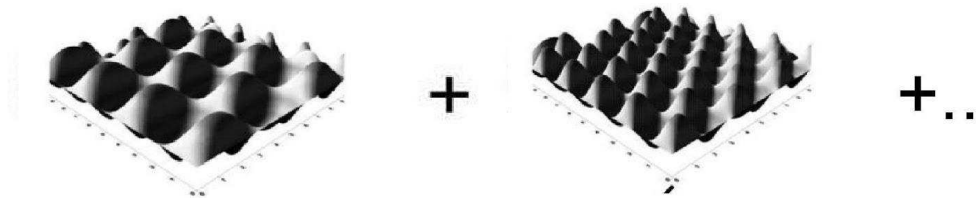 +..

Figure 13. *Pictorial view of of the 2D transform. Conversion of a map into a collection of regular surfaces.*

Adopting a more precise language, we start with a discreet map formed by MxN samples (remark that here parenthesis enclose Cartesian coordinates),

$$\delta k(0,0), \quad \delta k(0,1), \quad \delta k(0,2),\ldots\ldots\ldots, \delta k(m-1,n-1)$$

and we perform a two-dimensional Discrete Fourier Transform:

$$T_{f_x f_y} = \frac{1}{\sqrt{MN}} \sum_{x=0}^{M-1} \sum_{y=0}^{N-1} \delta k(x,y) \, e^{-i\theta_{f_x}} \, e^{-i\theta_{f_y}}$$

$$f_x = 0,\ldots, M-1 \qquad \theta_{f_x} = 2\pi \, x \, f_x / M$$

$$f_y = 0,\ldots, N-1 \qquad \theta_{f_y} = 2\pi \, y \, f_y / N$$

Notice that the complex exponentials $e^{-i\theta_{fx}}$ and $e^{-i\theta_{fy}}$ perform the above stated filtering over all the points $\delta k(x,y)$ of the map, so that the transform $T_{fxfy}$ for a given pair of frequencies $(f_x, f_y)$ is obtained. Thus, going over all the possible values of $f_x$ and $f_y$ the transform of all pairs of frequencies can be determined. Because complex exponentials are imaginary numbers—expressed as pairs of numerals assimilable, when plotted on an axis, to right-triangle legs—the magnitude of T$_{fxfy}$ is described by its modulus—the hypotenuse leg, as it were—computed as the positive squared root of the sum of the squares of the real and imaginary parts of the complex number; these are the weights or values we use in our spectra. For an overview on Fourier Transforms see Rahman (2011).

We can imagine the two-dimensional spectrum —see Fig. 15— as a surface displaying the relative values we need to assign to the pairs of frequencies in order to replay the original map—although we actually need two surfaces for that purpose, with the real and imaginary parts of the spectrum, they can be summarized in but one surface exposing the modulus, computed as described above. Furthermore, there is a symmetry among frequencies in regard to the distribution of weights. If we imagine the surface of the spectrum to be divided into four quadrants, designated as upper-outer quadrant, upper-inner quadrant, lower-inner quadrant and lower-outer quadrant, it turns out that the information in the lower-inner quarter is symmetrical to all quadrants. This notably reduces the analysis task — therefore a comparison can easily be drawn between a given series of spectra so as to find common features and also to disclose arising properties.

As can perhaps be deduced by now, the Fourier Transform works reverse-wise as well, in other words performing the Fourier Inverse Transform of the frequencies obtained from the map yields the map again, and inverse transforming a representative selection of frequencies produces a good approximation of it. The Fourier Inverse Transform may be viewed intuitively as reconstructing the original wave from the information contained in the complex numbers of its spectrum namely frequency and phase, the latter being the initial angle of a function at its origin.

Prior to studying the spectra we must see how to prepare the maps in order to perform the transform operation, and also how to prepare the spectra in order to compare them. This is explained in the next two subsections. We remark that in the following, the above-cited frequencies will be represented as follows: the pair of frequencies ($f_x$,$f_y$) is specified by (x-y), and so the pair ($f_2$,$f_3$) turns (2-3), for the sake of clarity.

## 5.2 Generation of equidistant nets

To perform the discreet version of the Fourier Transform we need maps where the values are ordered in an equidistant net. In order to assign a value of δk to each point (x,y) of the net, we use the inverse distance method. The algorithm applied is:

$$\delta k_{x,y} = \frac{\sum_{i=1}^{n} \frac{\delta k_i}{d_i^2}}{\sum_{i=1}^{n} \frac{1}{d_i^2}}$$

Where $\delta k_{x,y}$ is the value assigned to point (x,y) of the net, $\delta k_i$ is the residual of town i, $d_i$ is the distance between town i and point (x,y), and n is the extent of $\delta k_i$ values taken to calculate the value $\delta_{x,y}$. Adequate values of n and size of the net are empirically determined, and several tests have been made in order to choose its optimum mark. In order to minimize error and calculation time, a net size of 31x31 has been selected, together with a value of 10 for n.

## 5.3 Evolution of Spectra

Studying results across spectra implies a normalization so that frequencies from different sources be comparable. Thus let $p_i$ be a precise spectral power belonging to a given spectrum, we describe:

$$p'_i = p_i N/2S$$

where $S = \sum p_i$ and $N$ is the total of frequencies in the spectrum. The normalized powers $p'_i$ allow values from different spectra to be compared and, if seeking connection between maps and energy, to establish the correlation among spectral power and energy consumption (see Fig. 14). The frequencies whose power increases with energy can be conjectured to form an evolving structure, which is maintained by the expenditure of power coming through the exosomatic process. From this view alone, map evolution can be assessed, even with the assumption—provided the surface roughness—that frequency behaviour will be all but uniform—i.e. just a few frequencies will behave as desired. After developing the software, regarding Catalonia the maps between 1900 and 1970 have been considered, while concerning Galicia we tested those between 1900 and 1981. The good correlation obtained in Figures 3 and 4 for those opening periods is a good foundation upon which to start searching for evolving spatial structures.

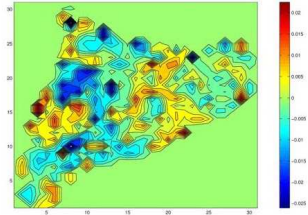
Catalonia 1900/1910

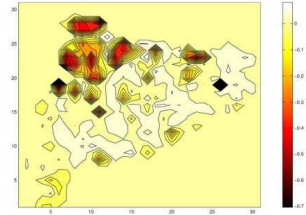
Catalonia 1960/1970

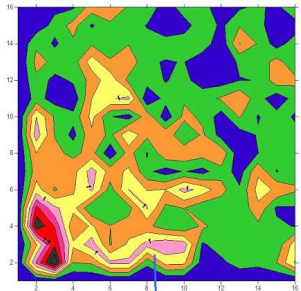

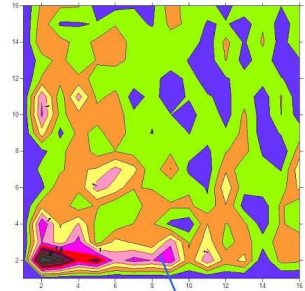

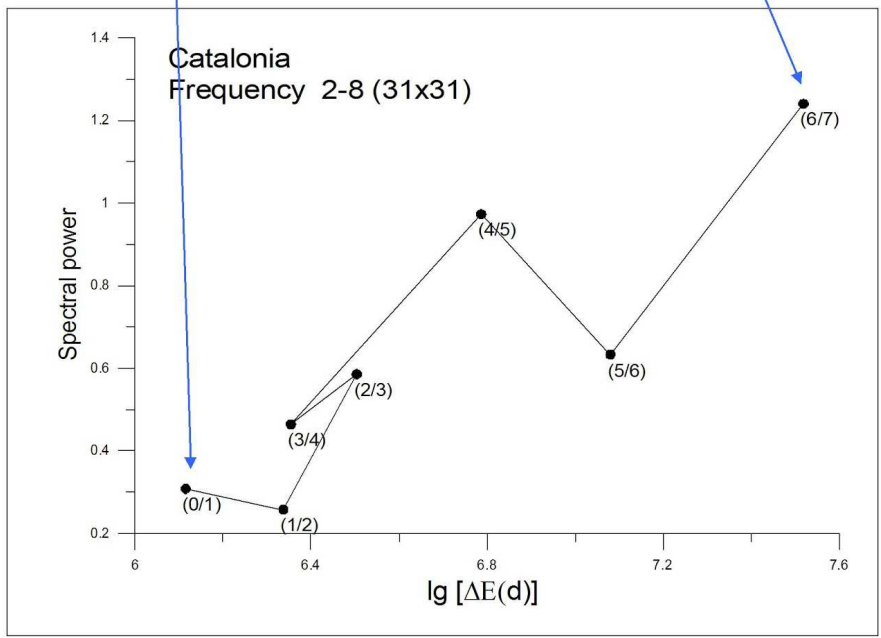

Figure 14. *Two maps related by the increase in weight of a given frequency. The maps on top emulate those in Figure 10, which once transformed provide the spectra in middle row images. Frequency (2-8) advances in power between the two considered spans of time, as displayed in bottom image. Detailed explanations about middle and bottom plots are given in Figures 15 and 16 respectively. The shading of figures is unrelated.*

## 5.3.1 Selecting energy-correlated frequencies in the case of Catalonia

From the obtained spectra for Catalonia (cf. Figure 15 for the case 1900/1910), we sketch the evolution of the normalized powers of each frequency with respect to the increase in energy consumption. As introduced above, the following figures point out how a given frequency varies when switching through the spectra, and how it correlates with energy. Here we select just the more important cases; in the case of Catalonia (Figure 16), the best result is shown by frequency (2-8) when working with a net of 31x31, which corresponds to a period (T) of (289 km - 41.3 km) according to :

$$T = L / (n-1)$$

with L = 289 km (side of the map's square), n = frequency. That is to say, if we imagine a sinusoidal surface repeating itself eight times in direction east-west, and two in direction north-south —this is what frequency (2-8) means—, it turns out that its weight (spectral power) in the different maps increases as more energy is used in the area. Recall that in virtue of the symmetry introduced in Section 5.1, considering the frequencies of the lower-inner quadrant is enough, hence reaching frequency (16-16) is suitable for a 31x31 net as plotted in Fig.15. Thus, spectral power of frequency (2-8) roughly scales with energy as $[[\Delta E(d)]^{0.63}/$antilogarithm

3.60], with [ΔE(d)] standing for increase in energy consumption and 0.63 being the line's slope in Figure 16. Value 3.60 is the independent term.

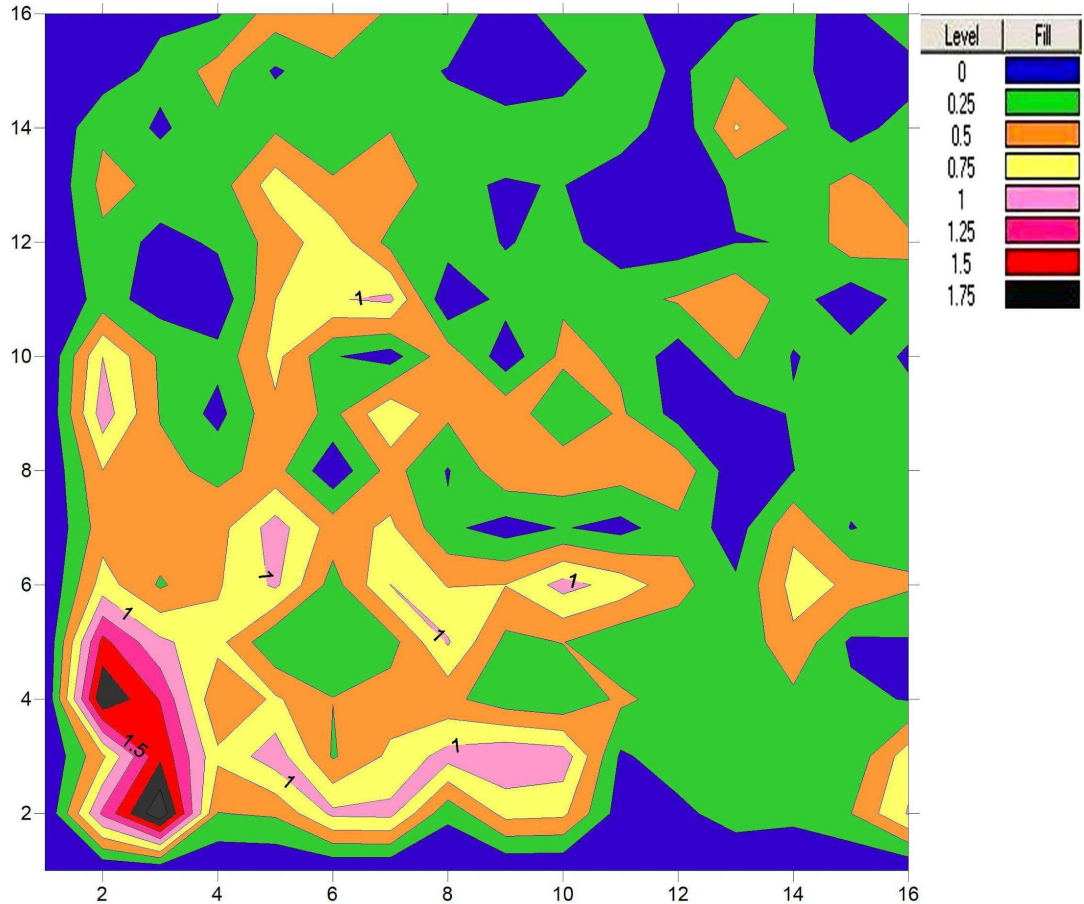

Figure 15. *The 1900/1910 spectrum for Catalonia. Its surface displays the relative values we need to assign to the pairs of frequencies in order to replay the original map. The values on both axis are the corresponding components of the frequencies—albeit the axes are inverted, i.e., power of frequency (2-8) is shown on point (8,2).*

Resuming to previous thermodynamic frame (cf. Sections 1 and 3), this result newly suggests the spontaneous appearance of a pattern in an environment where an exchange of energy and matter is taking place.

Likewise, being statistical in nature, this pattern does not precisely qualify as dissipative structure in the sense of Prigogine, but undoubtedly Figure 16 quantifies the emergence of a spatially-growing wave, linked somehow to system dynamics. Such a correlation is apparently unpublished in scientific literature, thus we cannot furnish a single example of similar data. Once more, given the absence of causal analysis involving the correlated variables, adequate interpretation remains a matter of debate.

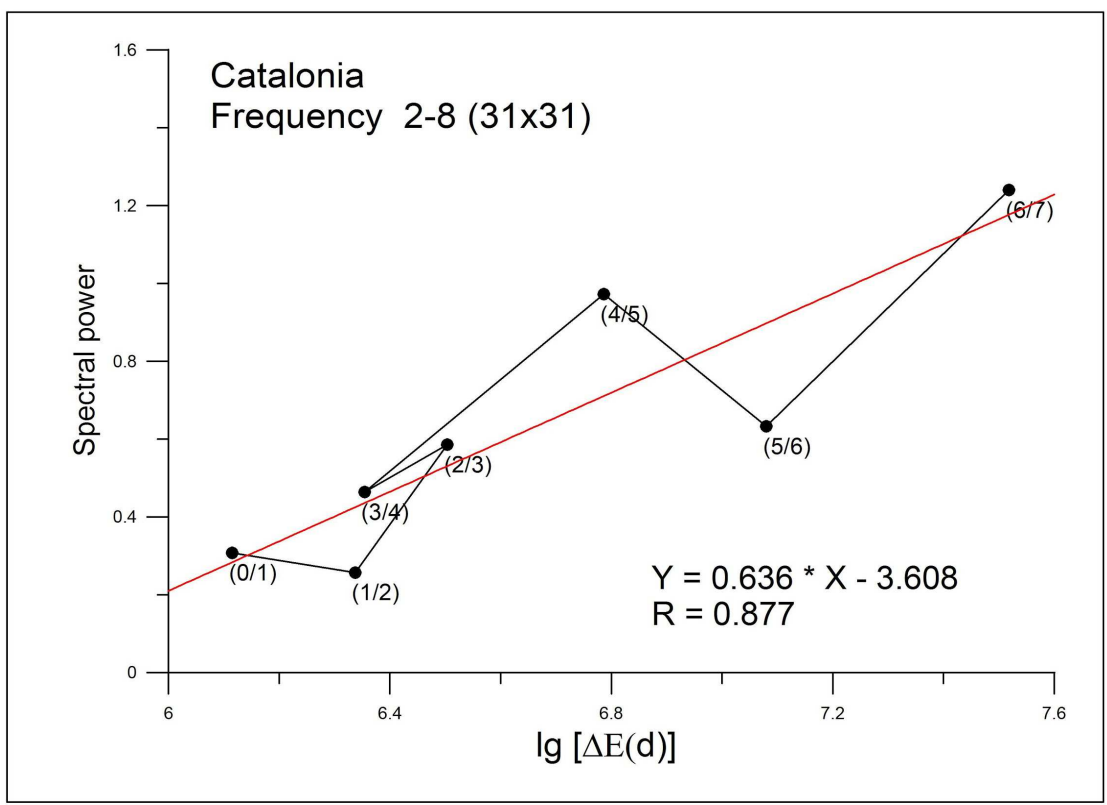

Figure 16. *Evolution of the normalized powers of frequency (2-8) for Catalonia with respect to the logarithm of the increase in energy consumption expressed in tce, termed lg[ΔE(d)]. The value in parenthesis is net size.*

We also get a quite correct behavior for frequency (2-4), but correlation is not maintained with respect to the last energy increase (Figure 17); here the importance lies in the fact that the involved power values are a wider fraction of the spectra, meaning that this frequency represents a more salient part of the map series. All and all, it must be remarked that either given frequency stands for a slight part of the cited spectra; for example frequency (2-8) adds up to 0.56 % of all the accounted weight, and frequency (2-4) reaches 1.4%. Let us reassert this point through examination of table 2.

| Table 2 Table of frequency weights [a] | | | | | | | |
|---|---|---|---|---|---|---|---|
| Catalonia | | | | Galicia | | | |
| $f_x$ | $f_y$ | relative weight | R | $f_x$ | $f_y$ | relative weight | R |
| 2 | 4 | 1,406 | 0,608 | 2 | 2 | 2,263 | 0,779 |
| 2 | 5 | 0,932 | 0,823 | 2 | 5 | 1,210 | 0,562 |
| 2 | 8 | 0,566 | 0,877 | 3 | 8 | 0,808 | 0,788 |
| 3 | 2 | 1,054 | 0,601 | 4 | 7 | 0,652 | 0,661 |
| 3 | 4 | 0,915 | 0,727 | 7 | 5 | 0,627 | 0,653 |
| 7 | 6 | 0,693 | 0,498 | 8 | 7 | 0,653 | 0,528 |
| 7 | 9 | 0,626 | 0,404 | 9 | 8 | 0,396 | 0,543 |
| 9 | 6 | 0,489 | 0,631 | | | | |
| | Total | 6,681 | | | Total | 6,610 | |

(a) Each weight is expressed as a percentage of the addition of weights regarding a given frequency, as against the total weight of the considered set of spectra. R stands for correlation coefficient calculated as in Fig. 16.

Although not shown here, frequency (2-5) also evolves quite well scoring similar to above values in terms of spectral power. In addition, frequencies (3-2) and (3-4) correlate well attaining closer values. Finally, (7-6), (7-9) and (9-6) can also be quoted even if achieving weaker powers. The whole set of cited frequencies amounts in this case to 6.7 % of the spectra; this value, if being a minor fraction of the total, further quantifies the sensibility of our approach. Seen in perspective, the reason for the exercise we are carrying out here is establishing additional evidence for energy linkage in population data; after the good correlation found in Section 3, the present results help further disentangle the energetic braid adding the spatial perspective to it all, yet their statistical weight is not as significant as that seen with the cited correlation.

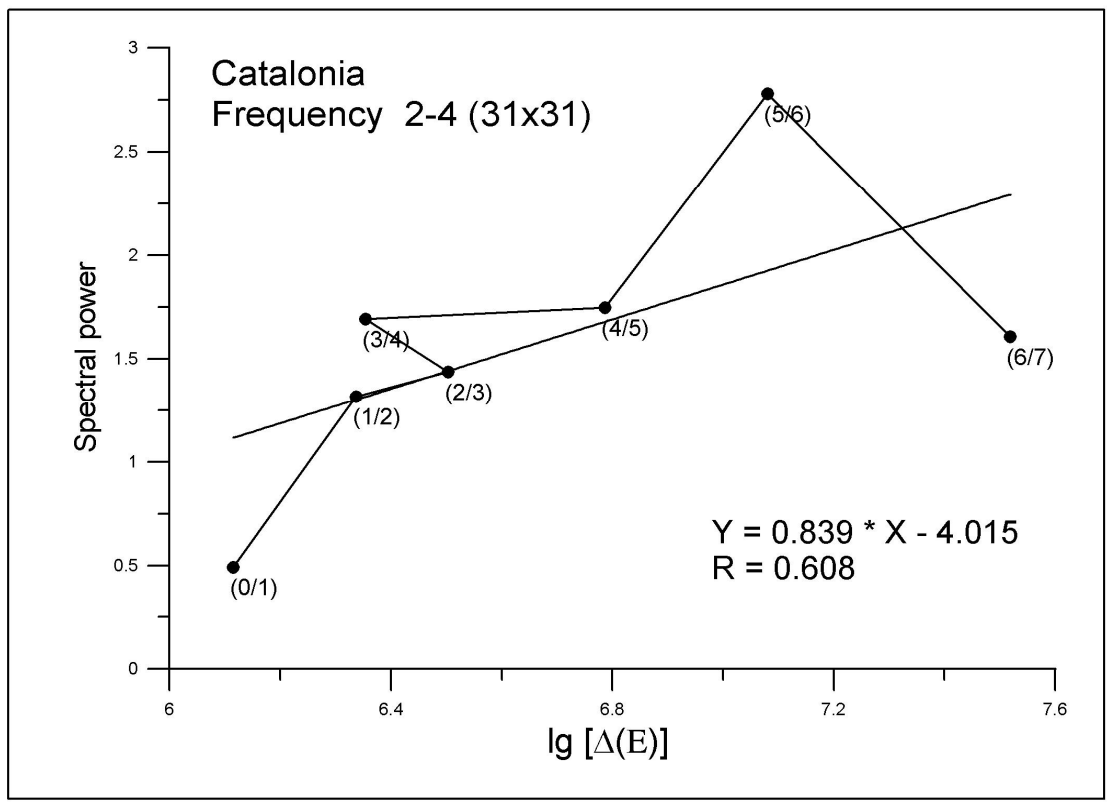

Figure 17. *Evolution of frequency (2-4) for Catalonia. The values on both axis emulate those in Figure 16 .*

As for analyzing recent years, the behavior of frequencies on current maps will be considered in further studies. In view of the obtained results we may tentatively perform some simulation exercises such as trend projections, so we expect reporting future work in this regard.

### 5.3.2. Spectra resulting from the maps of Galicia

As a matter of validation we can examine the series of spectra arising from the maps obtained in the case of Galicia. As we will see in the next figures, the major frequencies are different than the ones from Catalonia; indeed, the frequency which correlates better to the increase of energy consumption is (2-2) obtained working with a net of 31x31 (Figure 18); this frequency corresponds to a period of (227.3 - 227.3) km.  In the case of Catalonia this frequency shows different behaviour as pictured; at most one can say that the growing trend shown by Galicia is just observed regarding  the first three decades of the 20th century in Catalonia. Bearing in mind the more fully developed Catalan system, this insight could be interpreted to mean that Catalonia went through an initial development which in Galicia has not been achieved until the end of the century—this deciphering being presented with the admonition inherent to its speculative nature.

Although not displayed here, frequencies (2-5),  and (3-8) also show a certain correlation to energy, with spectral power values in the order of 2. Lastly, for powers next to 1 we can still quote frequencies  (4-7), (8-7), (7-5) and (9-8).  The entire group of described frequencies equals in this case to 6.6 % of the spectra; see the data presented in Table 2 for details.

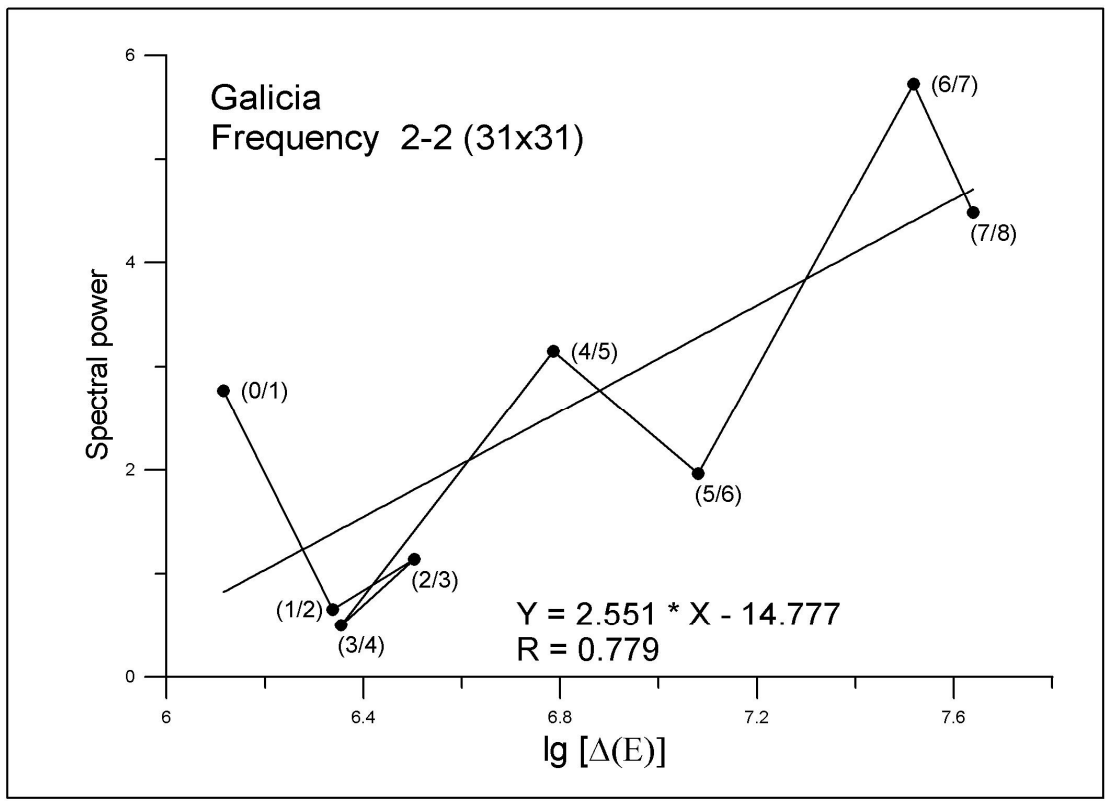

Figure 18. a)*Evolution of frequency (2-2) for Galicia.  The values on both*

*axis emulate those in Figure 16 .*

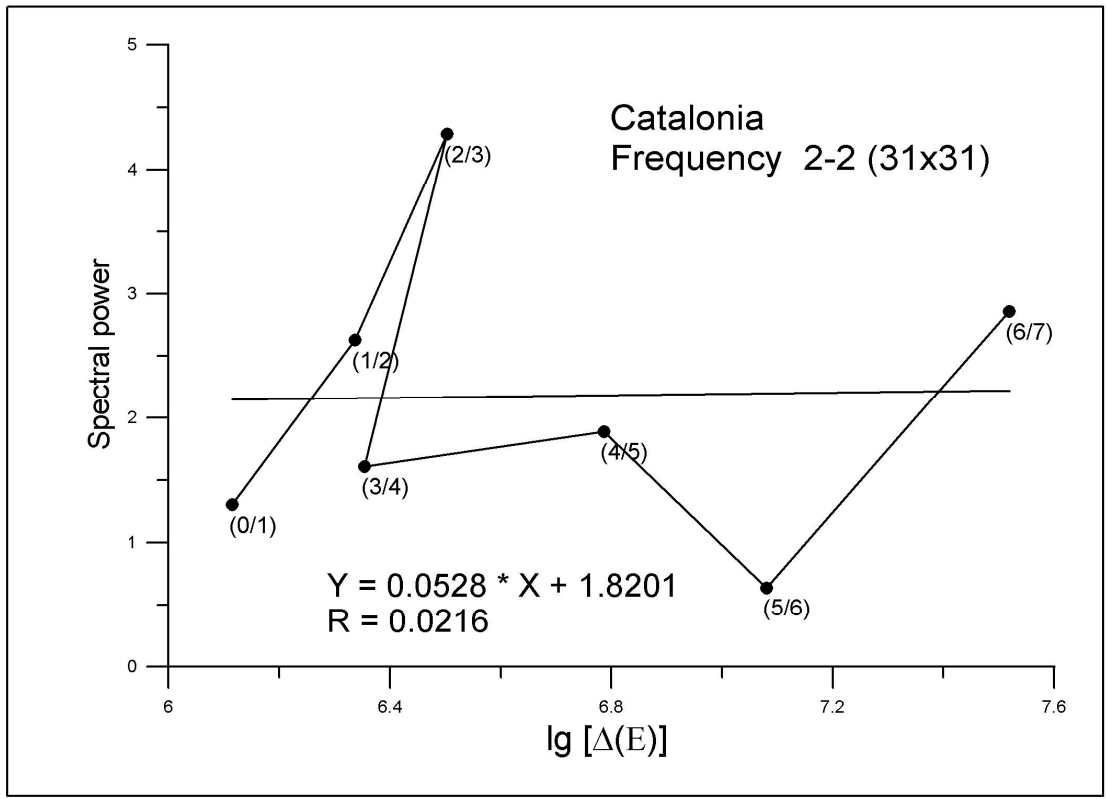

Figure 18. b) *Evolution of frequency (2-2) for Catalonia. The values on both axis emulate those in Figure 16 .*

# 6 Treatment and classification of the maps
## 6.1 Selection of maxima and evolution of error

Our next goal is identifying those points of the spectrum with relevant information in order to shape the original map, and this can be achieved through quantifying the faithfulness in reproducing the initial surface by a selection of frequencies, gradually evolving the process to meet the best fit. The associated rate of improvement can appear to be a convenient classification criterion, suitable to characterize the complexity of the maps as described in the next paragraphs.

For a given spectrum we accomplish our aim by means of tracking its whole extent in search for specially outstanding domains. Provided that the former is a square filled with the weights of the frequencies (cf. Fig 15), we can consider small squares to chase the whole area in quest of maxima. For example, if working with a square which side equals one, standing at frequency (2-2), we will add up the values corresponding to frequencies (2-2), (2-3), (3-2) and (3-3); whenever we get a score beyond a set bound value, the figures will be chosen. Otherwise, settled to be zero.

For the sake of clarity, if we envision the spectrum to be a surface, let's then consider two coordinate axes to hint at a given location, as the corresponding values $f_x$ and $f_y$ lead to the position of frequency ($f_x$-$f_y$). Assigned to this point is the value of the transform $T_{fx-fy}$, so the full spectrum includes all the transform values ranked by its matching pairs of frequencies. The course of above selection is carried out to select only the bumps i.e. maximum values of the spectrum, and the rest is set to zero. It is as if from a circus tent, we had chosen only the posts, the small tracking squares being the surface where to look for them. Furthermore we iterate the procedure through different values of the tracking square thus performing several selections. In so doing, a wide simplification is made; the square side is made to vary from 1 through 9, and for each selection the inverse Fourier transform of the chosen frequencies is computed so obtaining a series of approaches to the starting map.

We next calculate the error these approaches contain. Each point of the obtained map is compared to the original value and the difference squared and normalized dividing by the square of the grid; adding these values produces the error inherent to a given simplification. Hereby provided an original map we have a series of errors associated to a series of square sides. Figure 19 shows the case of decade 1900/1910 for Catalonia as an example; as a broad approach to the first results it can be said that error reduces as side of square increases and that comparing Catalonia with Galicia does not produce similar values until the second half of the century. Still, some duplicities may be contained in previous method— that is, when selecting the maxima some frequencies may get double-counted. To avoid this situation, a set of frequencies termed "net points" is created which we expose in next section.

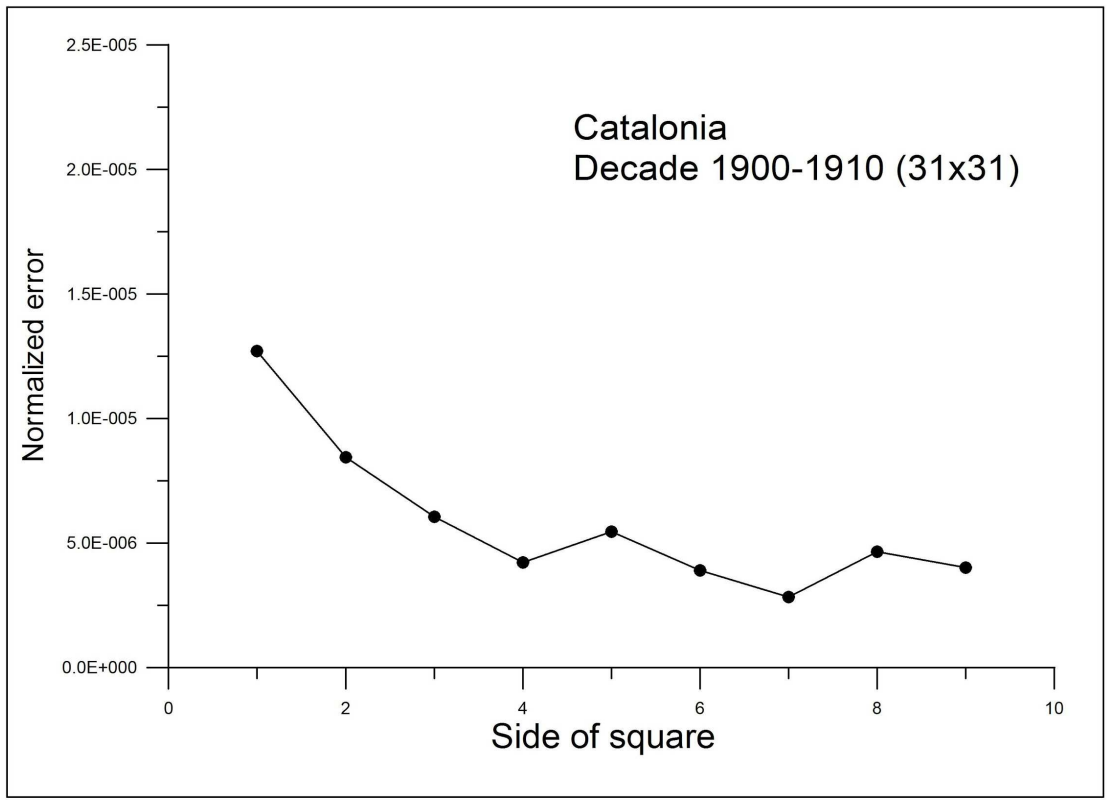

Figure 19. *Quantifying the faithfulness in reproducing the original surface. Provided an original map, we have a series of errors associated to spectrum-sampling squares which side is made to vary (see text). For calculating the error in any selection, the inverse Fourier transform is computed so obtaining an appreciation of the resemblance to the starting map. The graph plots normalized error values on Y-axis as a function of side of square values on X-axis in the case of Catalonia, concerning decade 1900/1910.*

## 6.2 A more precise criterion: net points

A filtering was performed avoiding repeated values of the spectrum thus producing a new simplified set to test with. Let's hence call "net points" this new array without duplicities and recalculate the contained error. In this round, once a map is settled along with a stated value of square side, there will be a given number of net points and a novel value of error with respect to original map. After iterating through different square values, a plotting of this error against net points is produced (see Fig. 20) and an exponential function of type $y=ae^{bx}$ is adjusted. In this case, y stands for normalized error and x for net points, while a and b are the parameters of the exponential function. Please note that symbols of this equation are incidentally introduced to wit unrelated to previously established variables.

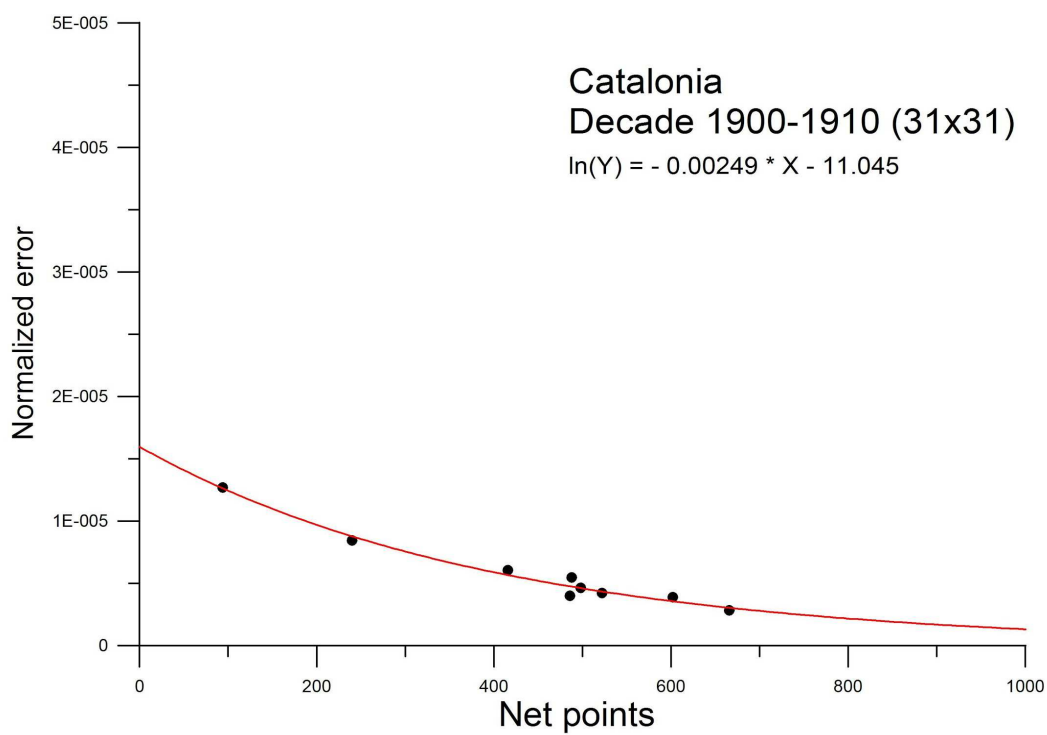

Figure 20. *Avoiding duplicities. A filtering was performed avoiding repeated values of the spectrum. Once a map is settled along with a stated value of square side, there will be a given number of "net points" and a novel value of error. The graph plots normalized error values on Y-axis,*

*as a function of "net points" values on X-axis for Catalonia regarding decade 1900/1910. Additionally, an exponential function is adjusted albeit expressed in a linearized form where ln(Y) is the Neperian logarithm of normalized error and X stands for net points. The slope b is nearly -0.0024 in this case, and parameter a is close to -11.044. Equation symbols are unrelated to previously established variables.*

Since our aim here is to quantify the complexity of maps, values of exponent b will be used as a criterion. The latter can be viewed geometrically as a slope which steepness accounts for the rate of improvement as more points are included. Hence lower values of it —in absolute value— hint at a lesser rate of recovery as more data are used in order to restore the original surface. The adjoint Table 3 which displays all adjusted exponents allows a first glimpse at relationships between values, and points to the need for a suitable criterion for pooling and comparison. Along this lines a likely evolution of the maps can be upholded.

| Table 3 | | |
|---|---|---|
| Table of adjusted exponents—or slopes b | | |
| Decade | Catalonia | Galicia |
| 1900-1910 | -0,002488 | -0,003404 |
| 1910-1920 | -0,003151 | -0,00346 |
| 1920-1930 | -0,002989 | -0,003877 |
| 1930-1940 | -0,002587 | -0,002642 |
| 1940-1950 | -0,003127 | -0,00257 |
| 1950-1960 | -0,002856 | -0,002549 |
| 1960-1970 | -0,002491 | -0,003478 |
| 1970-1981 |  | -0,003238 |

In order to get a suitable grouping, a plot can be produced for the adjusted functions normalizing as follows:

$$u = e^{bv}$$

where b stands for the above adjusted exponent. We identify v with "net points", and make it to range and produce a set of u values —assimilable to normalized error— through the above equation. Despite being theoretical, these values expose the trend defined by b thus allowing for proving through likeness. Again and except for b, presently introduced symbols must be treated as unrelated to earlier variables. The resulting plot allows of assorting exponents into classes, each of the latter exhibiting a very close behaviour of its elements as exemplified by figure 21. The adopted criterion implies that for a given class, when increasing the number of points, the improvement is tantamount.

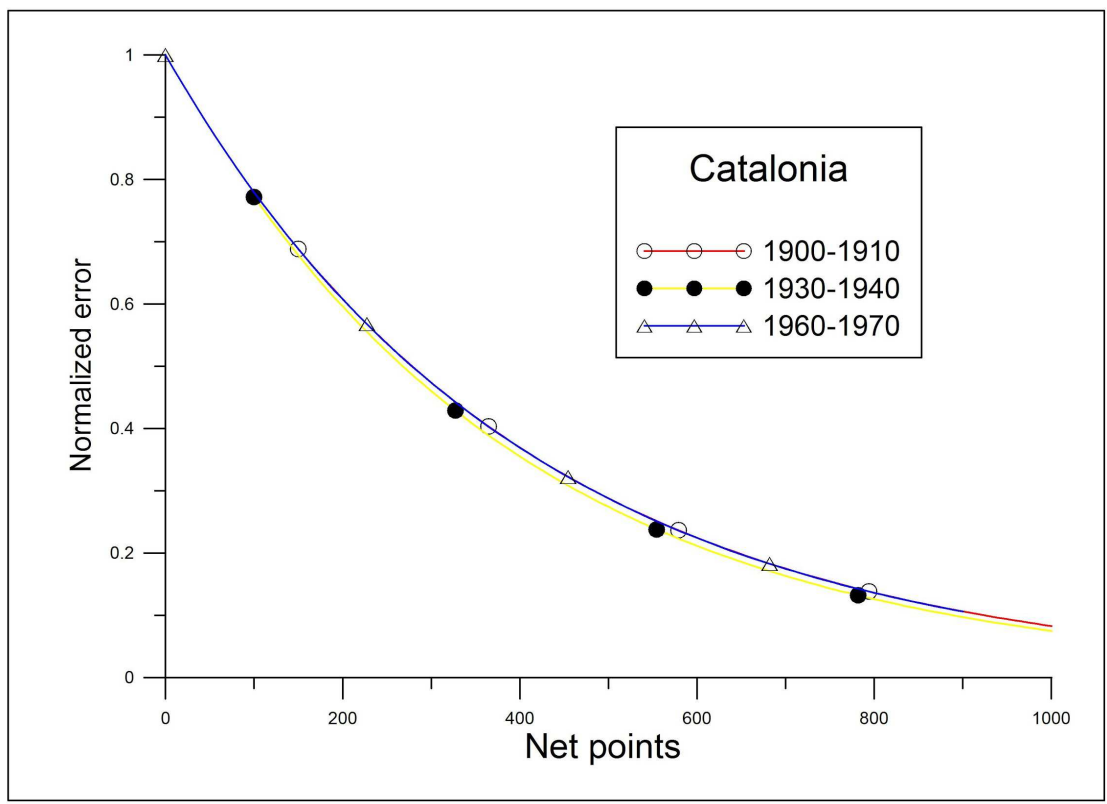

Figure 21. *Proving through likeness. Normalizing (see text) the values listed in Table 3 exposes the trend defined by b, thus favoring for grouping exponents into classes, each of the latter characterized by a very close pattern as exemplified here by Class A (see Table 4). The values on both axis emulate those in Figure 20, even if referred in the text as u —for normalized error—, and v —for net points— thus clarifying its theoretical origin .*

Grouping the maps into classes and labeling each one with the average value of b in absolute value, furnishes table 4. We underscore that the former averaging is feasible because of the very close match between elements of a class. From table 4, an abridged comparison of the maps belonging to the two zones is made possible. Hence, ordering the maps conforming to rising absolute value in b delivers the searched-for classification: A, F, C, B, D, E, and we are confronted with the next question, namely, "Why do the values of Galicia in years 1930, 1940 and 1950 resemble the ones obtained for Catalonia in years 1900, 1930 and 1960?" For an explanation, we can actually gauge evolution sketching the values in Table 3 as against time, thus producing Figure 22. We remark that on account of precision, original values of b are retrieved, even if referring to their absolute value.

| Table 4 Classes of maps | | |
|---|---|---|
| **Catalonia** | | |
| Class | Decades | Numerical mean of slope b in absolute value (x$10^{-3}$) |
| A | 0, 3, 6 | 2.522 |
| B | 1, 4 | 3.140 |
| C | 2, 5 | 2.922 |
| **Galicia** | | |
| Class | Decades | Numerical mean of slope b in absolute value (x$10^{-3}$) |
| D | 0, 1, 6, 7 | 3.396 |
| E | 2 | 3.877 |
| F | 3, 4, 5 | 2.587 |

The main result of present diagram is point (30-40) which is nearly shared by the two lines. Since the driving event of the epoch is Spanish Civil War, we are conducted to the conclusion that the demographic turmoil of the period induced a unification in the spatial features of the considered phenomenon, at least if treated in the manner just described.

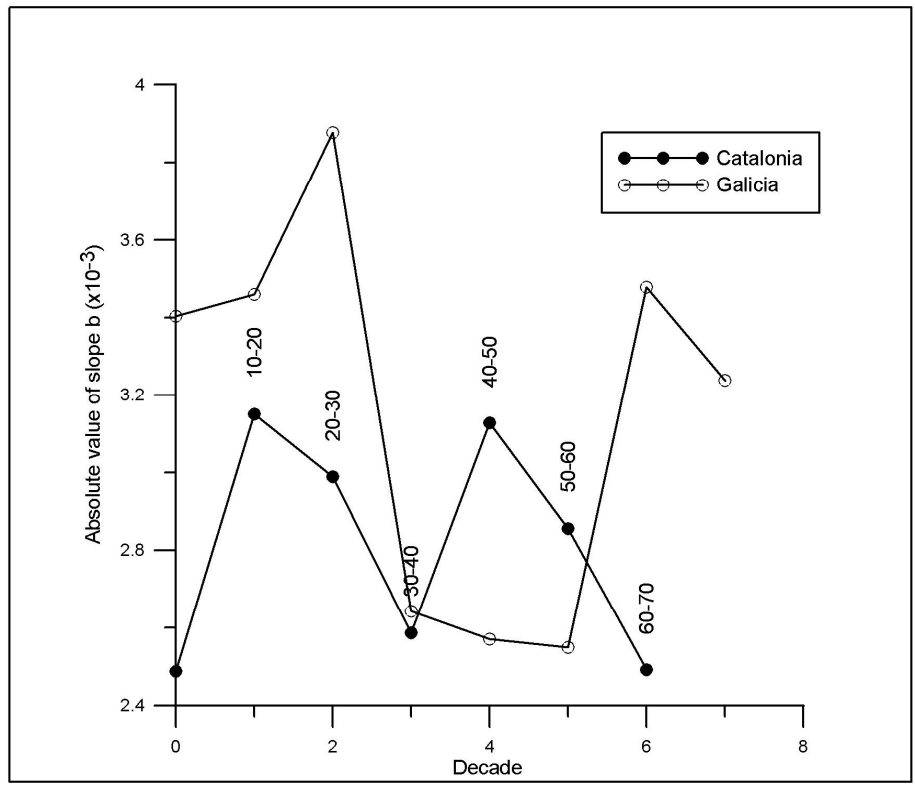

Figure 22. *Sketch of values in Table 3 as against time. Increasing the slope mark means that new structures are added as more spectral scope is provided thus ameliorating the restoring yield (see text). The values on the Y-axis are the absolute value of exponents or "slopes b" listed in Table 3, while X-axis features decades.*

The low count attained by the absolute value of the slope in this point invites to consider the redundancy of outstanding frequencies as a significant feature of the situation, for as it seems, the appearance of a performance gradient in map restoring implies the presence of critical neighbouring frequencies in the sampling area. To wit, if the maxima included in above tracking squares are accompanied by values casting the same structure, no performance gradient is detected and the absolute value of the slope sticks to a low score as bigger sampling area is considered; elseways new structures are added as more spectral scope is provided thus ameliorating the restoring yield and increasing the slope mark.

In addition to the exposed criteria, a hypothetical relationship can be assumed between arousal of new structures and emergence of socio economic development. As can be seen back in Figure 22 and following point (30-40), during the forties former values are retrieved in Catalonia, suggesting the precedence of a certain fabric which reacts quite rapidly once minimum conditions are met. Regarding Galicia, there is no recovery until the sixties thus associating, in this case, low absolute slope values to the underdevelopment that persisted in the area until the middle of the last century. On the other hand, Catalonia reaches low values again in the sixties despite economic growth occurring, therefore suggesting these values to point either toward or away from a lack of industrial pattern, in the latter case entailing to an incidental settlement once the exosomatic flow has been embodied.

Assuredly a classification has been achieved even though the credibility of the interpretation demands debate as some ambivalence remains hidden in the above results. Even so we guess that the topography of our maps has been typified through the methodology exposed herein, yet likewise we still studying other ways to relate them which we look forward to describing in ensuing reports.

## Conclusions

Our results suggest that, for the considered zones and periods, the rate of growth of human populations is not independent from the initial population but a function of it, the rate being greater along with a raising population. Such dependency depicts the emergence of autocatalytic effects owing to self organization in the system.

During a long stretch of 20th century and for the studied zones, the measured relationship between the municipalities' growth rates and their initial size correlates highly with commercial primary energy to a 0.9 coefficient. It is as if there was a granted distribution linked to every energy increase or as if the latter defined a migratory flow pattern. We interpret this as a system reorganization prompted by exosomatic energy flow, though the design of the study precludes a definitive causal analysis.

Regarding the considered zones and periods, energy and population are the factors that we denote as underlying in demographic distribution, thus allowing a model to be proposed. Displaying as a map the deviations from this model enables perceiving continuity in space, and recognizing an emerging pattern which structure was previously undetected. When considering positive outcomes, special population-clustering zones are detected, therefore emphasizing outstanding transport domains.

The obtained map collection may be studied in search for evolving spatial structures. Spectral analysis detects several frequencies which correlate well with energy; accordingly, the pattern resulting from the choice can be considered reliable for quantifying the raise of a spatially-growing wave.

The spectra of different maps can be simplified in such a way that allows measuring their relative complexity. In this way a study of the evolution of the latter can be performed, thus allowing establishment of some qualitative relations to demographic modifying factors.

# Acknowledgements

The ideas conducting to the development of the present project were suggested by the late Dr Ramon Margalef to whose memory this paper is dedicated.

Ministerio de Industria y Energía. 1961. *La energía en España 1945-1955*. Madrid.

Ministerio de Industria y Energía. 1978. *Evolución de los consumos provinciales en España 1960-1975*. Conferencia sobre la energía y recursos naturales. Madrid.

Odum H.T., Peterson N. 1996. Simulation and evaluation with energy systems blocks. *Ecological Modelling*, *93,* 155–173.

PCM. 1934. *Anuario Estadístico de España 1932-1933*. Presidencia del Consejo de ministros. D.G. del Inst. Geog. Catastral y de Estadística. Madrid.

Pearl, J. 2009. Causal inference in statistics: An overview. *Statist. Surv.* 3, 96--146. doi:10.1214/09-SS057. http://projecteuclid.org/euclid.ssu/1255440554., 28/8/2015.

Peschel M. and Mende W. 1986. *The Predator-Prey Model*. Springer.

PG. 1961. *Anuario Estadístico de España 1961*. Pres. del Gobierno, INE. Madrid.

Prigogine I. and Nicolis G. 1971. Biological order, structure and instabilities. *Quarterly Reviews of Biophysics,* 4, pp 107-148. doi:10.1017/S0033583500000615.

Pulselli R. M.,Pulselli F. M.,Ratti C. and Tiezzi E. 2005. Dissipative Structures for Understanding Cities: Resource Flows and Mobility Patterns. In: Boussabaine A.H.,Lewis J., Kirkham R.J.,and Jared G.E.M. (eds) *Proceedings of the 1st International Conference on Built Environment Complexity (Becon 2005)*. University of Liverpool

Rahman, M. 2011. *Applications of Fourier Transforms to Generalized*

*Functions*, WIT Press, ISBN 1845645642.